\documentclass[a4paper,11pt]{article}
\pdfoutput=1 

\usepackage{jcappub} 
\usepackage{graphicx,subfigure}

\usepackage[T1]{fontenc} 
\usepackage{color}

\hypersetup{
     colorlinks   = true,
     citecolor    = red,
     linkcolor = blue
     }
     
\newcommand\Tstrut{\rule{0pt}{1.5ex}} 
\newcommand\Tstrutbig{\rule{0pt}{2.6ex}}

\title{On the relationship between mean observations, spatial averages and the Dyer-Roeder approximation in Einstein-Straus models}
\author{S. M. Koksbang}
\affiliation{Department of Physics, University of Helsinki and Helsinki Institute of Physics, \\P.O. Box 64, FIN-00014 University of Helsinki, Finland}
\emailAdd{sofie.koksbang@helsinki.fi}


\abstract{
The redshift and redshift-distance relation in different Einstein-Straus models are considered. Specifically, the mean of these observables along 1000 light rays in different specific models are compared with predictions based on the Dyer-Roeder approximation and relations based on spatial averaging. It is shown that in certain limits, including those studied earlier in the literature, the Dyer-Roeder approximation and relations based on spatial averages agree with each other to a good precision regarding the redshift and redshift-distance relation and make good predictions of the mean of the exact relations. In limits where the two methods disagree, the Dyer-Roeder approximation clearly yields the better approximation of the true mean. This is explained by demonstrating the effect of boundary terms and integrated Sachs-Wolfe contributions but it is pointed out that the result seems to be valid for other Swiss-cheese models as well.
\newline\indent
An expression for the redshift drift in Einstein-Straus models is presented and used for studying the behavior of this quantity in particular Einstein-Straus models.
}

\begin{document}
	\maketitle
	\flushbottom
\section{Introduction}
The Universe is spatially inhomogeneous but we describe it and interpret observations with the spatially homogeneous Friedmann-Lemaitre-Robertson-Walker (FLRW) models. This works remarkably well as seen in e.g. \cite{Planck} (but is perhaps best illustrated by the precision required to identify the observational issues that standard cosmology is nonetheless facing - see e.g. \cite{H0-tension, Eleonora_mismatch,Eleonora_mismatch2,handley,obs_challenges}). While the observational issues with the $\Lambda$CDM model obviously lead to justified intrigue and raises important questions, an important issue is also raised by the good overall agreement between the FLRW model and observations made in a universe that is manifestly inhomogeneous, with observations based on light beams of different sizes and hence subjected to vastly different degrees of inhomogeneity. This issue is known as the fitting problem \cite{fittingproblem,fittingproblem2}.
\newline\newline
The fitting problem can be summarized as the questions of how, why and to what extent observations can be described by a single FLRW model and how this `best fit' FLRW model is related to the large scale geometry and dynamics  of the Universe (this `best fit' FLRW model will henceforth be referred to as the background). One approach to study the fitting problem is to consider how spacetime inhomogeneities affect observations. This has been done in a variety of ways including e.g. with perturbation theory (e.g. \cite{Kaiser, Marozzi1, Marozzi2, Marozzi3,cc1,cc2,Rose,Kibble,Bolejko_pertDR,syksy_pert}) and other types of analytical arguments (e.g. \cite{ZelDovich,DR1,DR2,Fleury,av_obs1,av_obs2,Linder,Linder_analytic_Study,Linder_alpha, misinterp, Futamase, sasaki, Coley1,Coley2}), using exact (e.g. \cite{Fleury, Fleuryetal,kantowski, scSZ1,scSZ2,scSZ3,scSZ5,scSZ6,scSZ7, scLTB1, scLTB2, scLTB3, scLTB4, scLTB5, scLTB6,scLTB7, scLTB8, scLTB9, scLTB10, scLTB11, scLTB12, scLTB13, scLTB14, scLTB15, scLTB16, scLTB17,lattice_exact}) and approximate (e.g. \cite{LW1,LW2,LW3,Larena, fitting_lattice}) cosmological models, different types of stochastic models (e.g. \cite{Mortell, Fleury_stochastic}), numerical relativity (e.g. \cite{startman,eloisa}) and Newtonian (e.g. \cite{Nbody1,Nbody2,Nbody3}) and relativistic \cite{lensing_gevolution} N-body simulations. The question has also been studied in light of cosmic backreaction \cite{fluid1,bc_fluidII}, in terms of how to describe mean observations in a universe with non-negligible cosmic backreaction (e.g. \cite{template1,template2,template3}) and in the timescape scenario \cite{timescape1,timescape2,timescape3,timescape4,timescape5,timescape6}. A recent review on the topic of small-scale inhomogeneities' effect on observations is given in \cite{Helbig}.
\newline\indent
One of the most well-known outcomes of the studies of how inhomogeneities affect observations is the Dyer-Roeder approximation \cite{DR1,DR2} (see also \cite{ZelDovich}). The Dyer-Roeder approximation is a redshift-distance relation which has been modified compared to the ordinary FLRW relation in an attempt to take into account the overall effect it has on the redshift-distance relation that thin beams of light (e.g. from supernovae) probe the Universe at scales so small that the corresponding matter distribution is best modeled (mainly) as opaque clumps. The approximation amounts to assuming that the redshift is unaffected by the inhomogeneities and hence is described according to the background FLRW model, while the angular diameter distance is approximated by neglecting the Weyl focusing and adjusting the Ricci focusing with a constant, $\alpha$. The resulting expression is ($c = 1$ throughout)
\begin{equation}\label{eq:DR}
	\frac{d^2D_A}{dz^2} + \left(\frac{2}{1+z}+\frac{d\ln H}{dz} \right) \frac{dD_A}{dz} = -\frac{4\pi G\rho}{H^2\left( 1+z\right)^2} \alpha D_A,
\end{equation} 
where $z$ is the redshift given according to the background FLRW model.
\newline\indent
The Dyer-Roeder expression has been modified in several works, e.g. with focus on a possible redshift dependence of $\alpha$ \cite{Linder_alpha, Teppo_alpha,Mortell}. This highlights the unfortunate detail that the value of $\alpha$ is not {\em a priori} predicted by the Dyer-Roeder approximation and must therefore either be considered a new model parameter or somehow estimated.
\newline\newline
The Dyer-Roeder approximation has been critiqued by several different authors. Critique has, for instance, been based on asserting that it is inconsistent to modify the mean density along the light ray without also modifying the expansion rate \cite{misinterp,cc3,EhlersSchneider}, including when computing the redshift\footnote{As in chapter 4 of \cite{LensingBook} one may note that multiplying $\rho$ by a constant factor leaves the fluid expansion rate unchanged according to the mass conservation law for dust. In \cite{LensingBook}, this is given as an argument for keeping the expansion rate fixed to that of the background in the Dyer-Roeder approximation. However, such argument seems to go beyond the original Dyer-Roeder approximation which solely addresses the question of mean observational relations and it is not clear how such mean observations can be combined with (local or average) dynamical equations - indeed, understanding how to do this is an essential part of the fitting problem.} (see also references in \cite{misinterp}). In addition, some studies show a disagreement with the Dyer-Roeder approximation. One interesting line of study which appears to not generally agree with the Dyer-Roeder approximation is that which is based on attempting to relate observations to spatial averages \cite{Linder,av_obs1,av_obs2}. Specifically, in \cite{av_obs1,av_obs2} it was assessed that if spacetime admits a foliation of spatial hypersurfaces of statistical homogeneity and isotropy with inhomogeneities evolving slowly compared to the time it takes a light ray to traverse the assumed homogeneity scale, then the redshift and angular diameter distance can be computed using spatially averaged quantities according to
\begin{equation}\label{eq:syksy}
\begin{split}
	\frac{d^2D_A}{dz_{\rm av}^2} + \left(\frac{2}{1+z_{\rm av}}+\frac{d\ln  H_{\rm av} }{dz_{\rm av}} \right) \frac{dD_A}{dz_{\rm av}} = -\frac{4\pi G\rho_{\rm av} }{H_{\rm av} ^2\left( 1+z_{\rm av}\right)^2} D_A,
	\end{split}
\end{equation}
where $\rho_{\rm av}$ is the smooth density obtained by averaging on some spatial domain above the homogeneity scale, and $H_{\rm av}$ is similarly a third of the averaged local expansion rate $\theta$. The redshift is here given by
\begin{equation}\label{eq:syksy_z}
	1+z_{\rm av} = \exp\left( \int_{t_e}^{t_0}dtH_{\rm av}\right) ,
\end{equation}
where $t_0$ is the time of observation and $t_e$ that of emission. The spatial average of a scalar $S$ is defined as $S_{\rm av}:=\frac{\int_D SdV}{\int_D dV}=:\left\langle S\right\rangle  $, where $dV$ is the proper infinitesimal volume element of the spatial hypersurfaces and $D$ is a spatial domain larger than the homogeneity scale.
\newline\indent
Equations \ref{eq:syksy} and \ref{eq:syksy_z} are based on the assumption that the content of the Universe can be described as dust that is comoving according to the foliation with statistically homogeneous and isotropic spatial hypersurfaces. The equations were generalized in \cite{av_obs2} but this generalization will not be needed here.
\newline\indent
The predicted relations in equations \ref{eq:syksy} and \ref{eq:syksy_z} will in the following be referred to as the ``relations based on spatial averages''.
\newline\newline
Both the Dyer-Roeder approximation and the relations based on spatial averages are meant to describe the redshift-distance relation in a mean sense, i.e. after averaging over a large number of light rays so that fluctuations along the individual light rays due to local inhomogeneities can be neglected. While the Dyer-Roeder approximation explicitly assumes that there are opaque regions that light rays cannot sample, the relations based on spatial averages were derived assuming that light rays trace spacetime ``fairly'' i.e. that there are no opaque regions. A naive extension of the analyses in \cite{av_obs1,av_obs2} to include opaque regions would, however, imply that the averages in equations \ref{eq:syksy} and \ref{eq:syksy_z} should simply only include the regions that light rays are permitted to travel through. One then sees that this naive extension and the Dyer-Roeder approximation do not agree since the Dyer-Roeder approximation misses the effect inhomogeneities have on the redshift.
\newline\newline
The Dyer-Roeder approximation is largely an assumption and e.g. does not provide a recipe for computing $\alpha$, making it somewhat less useful for making predictions. On the other hand, the method based on spatial averages is based on a thorough, detailed analysis which, to this author's knowledge, has not been critiqued. It is therefore intriguing that a study based on an exact solution to the Einstein equations (Einstein-Straus models \cite{EisteinStraus1}) showed that for the studied class of cosmological solutions, the Dyer-Roeder approximation is actually correct \cite{Fleury}. The study did not include much consideration of spatial averages so it did not pinpoint why the analyses of \cite{av_obs1,av_obs2} as well as the earlier work in e.g. \cite{Linder} seems to fail for spacetimes with opaque regions.
\newline\newline
The literature does not yet contain clear and hence satisfying answers to the questions collectively going under the name the fitting problem. Obtaining these answers is vital for fully understanding e.g. the significance of parameter determinations obtained when interpreting observations with FLRW models, including possible biases which, if not taken into account, could lead to e.g. wrong determinations of global energy densities, curvature and expansion rate. The goal with the study presented here is therefore to supplement the study in \cite{Fleury} by considering light propagation in Einstein-Straus models from the perspective of spatial averages, in order to obtain a better understanding of under what circumstances the two redshift-distance relations in equations \ref{eq:DR} and \ref{eq:syksy} are applicable.
\newline\newline
While the redshift and redshift-distance relation are important for interpreting many well-known observables such as CMB, supernovae and BAO observations, future astrophysical surveys will include new types of observables. One such example is the redshift drift, $\Delta z$, which describes how the observed redshift of a given comoving source is seen to change with time (according to a comoving observer). In an FLRW universe, the redshift drift in a time interval $\Delta t_0$ is given by
\begin{equation}
\Delta z = \Delta t_0 (1+z)\left( a_{,t}(t_0) - a_{,t}(t_e)\right) ,
\end{equation}
i.e. the redshift drift is non-vanishing if the expansion rate of the Universe changes with time and its sign depends on whether the expansion accelerates or decelerates between the times of observation and emission.
\newline\indent
In an inhomogeneous universe, the expression for the redshift drift is necessarily more complicated. The expression has been identified in spherically symmetric models with an observer placed in the center of symmetry \cite{Clarkson_zdrift}, Lemaitre-Tolman-Bondi models \cite{LTB_zdrift}, Szekeres models \cite{Sz_zdrift1,Sz_zdrift2,Sz_zdrift3}, Stephani models \cite{dz_Stephani} and Bianchi I models \cite{dz_bianchi}. A fairly simple expression for the redshift drift in a general spacetime can be obtained quite easily by using the exact definition, $\Delta z :=\frac{dz}{dt_0}\Delta t_0$, at least for the case of comoving source and observer. This is shown in \cite{digselv_manuscript} where it is also shown by explicit examples that the mean redshift drift in inhomogeneous cosmological models may deviate from the drift of the mean redshift (in agreement with the results in \cite{dz_digselv2}). The method presented in \cite{digselv_manuscript} will here be used to study the redshift drift in Einstein-Straus models.
\newline\newline
In section \ref{sec:modelSetup} below, the Einstein-Straus models will be introduced. Section \ref{sec:lightpropagation_einsteinStraus} describes how light propagation can be described in the models. Numerical results and analytical investigations of the (mean) redshift, redshift-distance relation and redshift drift are presented in section \ref{sec:results} and a summary with concluding remarks is given in section \ref{sec:Summary}.

\section{The Einstein-Straus model}\label{sec:modelSetup}
Inhomogeneous cosmological models can be constructed as so-called Swiss-cheese models where spatially spherical regions of an FLRW spacetime are replaced by other solutions to the Einstein equations. The Einstein-Straus model was introduced in \cite{EisteinStraus1} as the first ever Swiss-cheese model constructed by replacing spatially spherically symmetric patches of an Einstein-de Sitter (EdS) spacetime with spherically symmetric exterior Schwarzschild patches. Light propagation was studied in these models already in 1969 in \cite{kantowski} and was recently revisited in \cite{Fleury,Fleuryetal} where the model was extended to include a cosmological constant.
\newline\newline
The Swiss-cheese model studied here is similar to the original Einstein-Straus model so the line element of the cheese (the FLRW ``background'') is that of the EdS model, i.e.
\begin{equation}
	ds^{\text{cheese}} = -dT^2 + a^2\left( d\xi^2 + \xi^2d\Omega^2 \right) ,
\end{equation}
where $a = \left( \frac{T}{T_0}\right)^{2/3}$ is the scale factor normalized to 1 at present time, $T = T_0$.
\newline\newline
The line element of the Schwarzschild ``holes'' is given by
\begin{equation}
ds^{\text{hole}} = -A(r)dt^2 + B(r)dr^2 + r^2d\Omega^2.
\end{equation}
For the exterior Schwarzschild solution, the metric functions $A$ and $B$ are given by $A = 1-\frac{2M}{r}$, $B = \left( 1-\frac{2M}{r}\right) ^{-1}$, where $M$ is the effective gravitational mass of the central/interior part of the Schwarzschild spacetime. The exterior Schwarzschild metric represents the vacuum outside a spherically symmetric massive body and it can be combined with other spacetimes representing this interior massive region. The interior region can for instance be another FLRW model or a spherically symmetric Stephani solution \cite{Krasinski_bog} but the simplest interior region is given by the interior Schwarzschild solution. In this case, the metric functions are given by
\begin{equation}
A(r) = \left\{ \begin{array}{rl}
\frac{1}{4}\left(  3\sqrt{1-\frac{2M}{r_{\rm bie}}} - \sqrt{1-\frac{2Mr^2}{r_{\rm bie}^3}} \right) ^2  &\text{if} \,\, r<r_{\rm bie} \\
1-\frac{2M}{r} &\mbox{ otherwise}
\end{array} \right.
\end{equation}
and
\begin{equation}
B(r) = \left\{ \begin{array}{rl}
\left( 1-\frac{2Mr^2}{r_{\rm bie}^3}\right) ^{-1} &\text{if} \,\, r<r_{\rm bie} \\
\left( 1-\frac{2M}{r}\right) ^{-1} &\mbox{ otherwise}
\end{array} \right.,
\end{equation}
where $r_{\rm bie}$ denotes the value of $r$ on the boundary between the interior and exterior Schwarzschild metric. An interior Schwarzschild solution is usually not considered in cosmological contexts because it describes a perfect fluid with a (in general) non-physical equation of state parameter. However, the principles of light propagation do not depend on equation of state parameters being physically reasonable. Hence, the interior Schwarzschild region may be used for studying the principles of light propagation in situations where an interior region is desired. Minor results will therefore be presented based on including the interior Schwarzschild regions in the Einstein-Straus model.
\newline\indent
 The interior Schwarzschild solution has a constant density related to $M$ by $\rho_{\text{interior}} = \frac{M}{\frac{4}{3}\pi r_{\rm bie}^3}$ and an $r$-dependent pressure \cite{blaaGRbog}
\begin{equation}
p = \rho\frac{\sqrt{1-\frac{2Mr^2}{r_{\rm bie}^3} } - \sqrt{1-\frac{2M}{r_{\rm bie}}} }{3\sqrt{1-\frac{2M}{r_{\rm bie}}} - \sqrt{1-\frac{2Mr^2}{r_{\rm bie}^3}}}.
\end{equation}
This form of the pressure is determined by requiring that the pressure vanishes on the boundary between the interior and exterior Schwarzschild regions. For the models studied here, the pressure is subdominant to the density to the extent that it is negligible for the precision needed here.
\newline\newline
In order to be an exact solution to the Einstein equations, a Swiss-cheese model must fulfill the Darmois junction conditions \cite{Darmois} which require the extrinsic curvature as well as the metric on the boundary between different solutions to be continuous. The Darmois junction conditions for the Einstein-Straus model are well known (see e.g.  \cite{Fleuryetal, Ishak_kottler}) and are 
\begin{equation}
	r_{\rm b} = a\xi_{\rm b},
\end{equation}
where $r_{\rm b},\xi_{\rm b}$ are the values of $r,\xi$ at the junction of the two spacetimes, and
\begin{equation}
	 M = \frac{4\pi G \rho_0}{3}\xi_{\rm b}^3,
\end{equation} 
where $G$ is Newton's constant and $\rho_0$ is the present time energy density in the EdS model. For a given $M$, $\xi_{\rm b}$ (and hence $r_{\rm b}$) is uniquely determined. On the other hand, $r_{\rm bie}$ can be chosen freely, though keeping in mind that $\frac{GM}{r_{\rm bie}}<\frac{4}{9}$ is required to avoid infinite pressure \cite{blaaGRbog}, and that the model can only be considered for time intervals where $r_{\rm bie}<r_{\rm b} = r_{\rm b}(t)$.
\newline\indent
Models with three different sets of parameter values will be considered here, with model parameters specified in section \ref{subsec:implementation}

\section{Light propagation formalism for the Einstein-Straus model}\label{sec:lightpropagation_einsteinStraus}
The straightforward way to study light propagation in an Einstein-Straus model is to use the null-geodesic equations
\begin{equation}
\frac{d}{d\lambda}\left(g_{\alpha\beta}k^{\beta} \right)  = \frac{1}{2}g_{\mu\nu,\alpha}k^{\mu}k^{\nu} 
\end{equation}
and the transport equation (see e.g. \cite{arb_space})
\begin{equation}
\frac{d^2D^a_b}{d\lambda^2} = T^a_c D^c_b,
\end{equation}
where $\lambda$ is an affine parameter along the null geodesic and  $T^a_c$ is the tidal matrix which can be written as
\begin{equation}
T_{ab} = 
\begin{pmatrix} \mathbf{R}- Re(\mathbf{F}) & Im(\mathbf{F}) \\ Im(\mathbf{F}) & \mathbf{R}+ Re(\mathbf{F})  \end{pmatrix} ,
\end{equation}
where $\mathbf{R}: = -\frac{1}{2}R_{\mu\nu}k^{\mu}k^{\nu}$ and $\mathbf{F}:=-\frac{1}{2}R_{\alpha\beta\mu\nu}(\epsilon^*)^{\alpha}k^{\beta}(\epsilon^*)^{\mu}k^{\nu}$. $R_{\alpha\beta\mu\nu}$ is the Riemann tensor and $\epsilon^{\mu} := E_1^{\mu} + iE_2^{\mu}$ is a complex combination of two orthogonal vectors spanning the space orthogonal to both $k^{\mu}$ and $u^{\mu}$. $E_1^{\mu}$ and $E_2^{\mu}$ must fulfill certain orhonormality conditions, namely $E_i^{\mu}E^j_{\mu} = \delta^i_j$, $E_i^{\mu}u_{\mu} = 0 = E_i^{\mu}k_{\mu}$, $i,j\in{1,2}$. $E_1^{\mu}$ and $E_2^{\mu}$ are typically parallel propagated along the light rays with initial conditions fulfilling these requirements. When the (comoving) observer is placed in the FLRW cheese, initial conditions can be chosen as
\begin{equation}
	E^{\mu}_1\propto \left( 0,\xi^2\left( \left( k^{\theta}\right) ^2 + \sin^2(\theta)\left( k^{\phi}\right) ^2\right)  , -k^rk^{\theta}, -k^rk^{\phi}\right) 
\end{equation}
\begin{equation}
	E^{\mu}_2 \propto \left(0,0,k^{\phi}, \frac{-k^{\theta}}{\sin^2(\theta)}\right) .
\end{equation}
The Weyl tensor vanishes in the EdS region so $k^{\mu}k_{\mu} = 0$ implies that $\mathbf{F}= 0$ there. The exterior Schwarzschild region is a vacuum spacetime so $\mathbf{R} = 0$ but $\mathbf{F}\neq 0$. In the interior Schwarzschild spacetime region, both $\mathbf{F}$ and $\mathbf{R}$ are non-vanishing.
\newline\indent
The angular diameter distance is given by $D_A = \sqrt{|\det D|}$. Therefore, the transport equation is solved simultaneously with the geodesic equations in order to obtain the exact redshift-distance relation along individual light rays.

\subsection{Redshift drift in Einstein-Straus models}
The redshift drift in Einstein-Straus models can be computed following the procedure presented in \cite{digselv_manuscript}, i.e. by differentiating $1+z = \frac{\left(u^{\alpha}k_{\alpha} \right) _e}{\left(u^{\beta}k_{\beta} \right) _0}$ with respect to observer present time. The expression depends on the position of observer and emitter. Here, the redshift drift will only be computed with both observer and emitter in the FLRW region. In this case, the expression for the redshift drift is
\begin{align}
\begin{split}
\frac{\Delta z}{\Delta T_0} &= -\frac{(1+z)}{k^T_0}\frac{\partial k^T}{\partial T}|_0 + \frac{1}{k^T_0}\frac{1}{1+z}\frac{\partial k^T}{\partial T}|_e
\\ &= -\frac{(1+z)}{\left( k^T_0\right) ^2}\left(\frac{dk^T}{d\lambda}-k^ik^{T}_{,i} \right)|_0  + \frac{1}{ k^T_ek^T_0}\frac{1}{1+z}\left(\frac{dk^T}{d\lambda}-k^ik^{T}_{,i} \right)|_e
.
\end{split}
\end{align}
In the limit where the light ray only travels in the EdS region, $k^T = \frac{-1}{a}$ and $k^T_{,T} = \frac{1}{k^T}\frac{dk^T}{d\lambda}$, and the above expression reduces to the ordinary FLRW expression $\Delta z = \Delta T_0\left(H_0(1+z) -H_e \right) $ as it should. However, in the Einstein-Straus model, $\frac{dk^T}{d\lambda}:=k^{\alpha}k^{T}_{,\alpha}$ will not necessarily reduce simply to $k^{T}k^{T}_{,T}$ along the entire light ray. The redshift drift may therefore differ from $\Delta z_{\rm EdS}$ if the particular light ray traverses one or more Schwarzschild regions.
\newline\newline
In order to compute the redshift drift, $k^T_{,T}:=\frac{\partial k^T}{\partial T}$ must be computed along the light rays. This is achieved by solving the equations $\frac{dk^{\mu}_{,\nu}}{d\lambda}$ simultaneously with the geodesic equations and the transport equations. The expressions for $\frac{dk^{\mu}_{,\nu}}{d\lambda}$ are found from the relation   \cite{scSZ7}
\begin{equation}
	\frac{d}{d\lambda}k^{\mu}_{,\nu} = \frac{\partial }{\partial x^\nu}\frac{dk^{\mu}}{d\lambda}-k^{\beta}_{,\nu}k^{\mu}_{,\beta}.
\end{equation}
Since the observer is placed in the FLRW region, initial conditions can be set according to a spatially homogeneous universe where $k^{\mu}_{,x}=k^{\mu}_{,y} = k^{\mu}_{,z} = 0$ (which becomes somewhat more complicated when transforming to spherical coordinates which were used here). 
\newline\newline
Before moving on, note that the drift of $z_{\rm av}$ is given by a formula similar to that of FLRW models, namely as $\Delta z_{\rm av} = \Delta T_0\left((1+z_{\rm av})H_{0,av}-H_{\rm av} \right) $ (see e.g. \cite{dz_digselv1}).

\subsection{Numerical implementation}\label{subsec:implementation}
\begin{figure*}
	\centering
		\includegraphics[scale = 0.60]{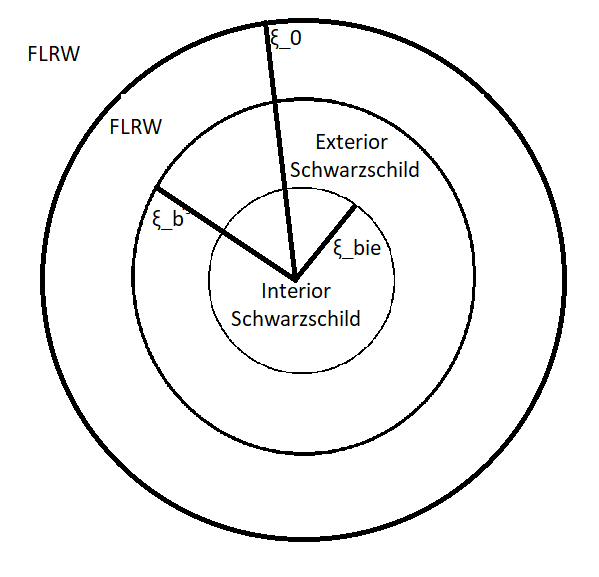}
	\caption{Sketch of Swiss-cheese setup illustrating the relationship between $\xi_0$, $\xi_{\rm bie}$ and $\xi_{\rm b}$. The coordinate $\xi_{\rm bie}:=ar_{\rm bie}$ is defined analogously to $a\xi_{\rm b}= r_{\rm b}$.}
	\label{fig:sketch}
\end{figure*}
The redshift-distance relation is computed by assembling the Einstein-Straus models on-the-fly, i.e. along the given light ray during its propagation. In practice, each light ray has been initialized in the EdS background at a fixed comoving value of $\xi$, $\xi = \xi_0$ (the value of $\xi_0$ for each studied model is given in table \ref{table:models}). The initial values of $k^r, k^{\theta}$ and $k^{\phi}$ are random, though with $k^r\leq0$ and such that the null condition is fulfilled with the initial condition $k^T=-\frac{1}{c}$. When a light ray again reaches $\xi = \xi_0$, it is turned back towards the Schwarzschild structure with a new random impact parameter. If the inner Schwarzschild region is considered opaque, a light ray reaching $r\leq r_{\rm bie}$ is moved back to the previous time it reached $\xi = \xi_0$ and turned back towards the Schwarzschild structure with new (random) impact parameter.
\newline\indent
A sketch of the setup with boundaries at three radial coordinate values is shown in figure \ref{fig:sketch}. Three different models will be considered, with parameter values given in table\ref{table:models}.
\newline\newline

\begin{table}
	\centering
	\begin{tabular}{c c c c c}
		\hline\hline\Tstrutbig
		Model name & $M$ ($M_{\odot}$) & $r_{\rm bie}$ (Mpc)& $\xi_{\rm b}$ (Mpc)& $\xi_0$ (Mpc)\\
		\hline\Tstrutbig
		model 1& $10^{15}$  & 1 & 12.06& 12.3 \\
		model 2&  $10^{16}$ & 10 & 25.99 & 26.1 \\
		model 3&  $10^{16}$ & 10 & 25.99& 28.5 \\
		\hline
	\end{tabular}
	\caption{Model parameters of the studied Einstein-Straus models. }
	\label{table:models}
\end{table}

At the boundary between the EdS and Schwarzschild region, a coordinate transformation is necessary. Specifically, $(T,\xi)$ are transformed into $(t,r)$ and the null-vector and screen space basis vectors are also transformed accordingly. The transformation rules were given in \cite{Fleuryetal} and are here re-stated for convenience:
\begin{equation}
	\frac{dT}{dt} = A
\end{equation}
\begin{equation}
		r = a\xi
\end{equation}
\begin{equation}\label{eq:kr_trans}
		k^r = ak^\xi + \sqrt{1-A}k^T
\end{equation}
\begin{equation}\label{eq:kt_trans}
	k^t = \frac{1}{A}k^T + \frac{a}{A}\sqrt{1-A}k^{\xi},
\end{equation}
(Note that for the particular models studied here, $|A(r_{\rm b}(T))-1|<10^{-4}$ so setting $t = T$ on the boundary is a reasonable approximation.)
\newline\newline
The partial differentials of the FLRW and Schwarzschild coordinates $\left( \frac{\partial T}{\partial r}  \text{ etc.} \right)$ can be seen by comparing equations \ref{eq:kr_trans} and \ref{eq:kt_trans} with the general formula for coordinate transformations of 4-vectors, $k^{\mu'} = \frac{\partial x^{\mu'}}{\partial x^{\nu}}k^{\nu}$. These can then be used to transform covariant derivatives of $k^{\mu}$ between the two coordinate systems according to $k^{\mu'}_{;\nu'} = \frac{\partial x^{\mu'}}{\partial x^{\mu}}\frac{\partial x^{\nu}}{\partial x^{\nu'}}k^{\mu}_{;\nu}$ and from these, the partial derivatives, $k^{\mu}_{,\nu} = k^{\mu}_{;\nu} - \Gamma^{\mu}_{\nu\gamma}k^{\gamma}$, can be computed when going between FLRW and Schwarzschild regions.

\section{Mean observations in Einstein-Straus models}\label{sec:results}
This section presents the results from propagating 1000 light rays out to $z = 1$ in different Einstein-Straus models and compares the results to the Dyer-Roeder approximation and the relations based on spatial averages. An individual light ray may be in either region type (Schwarzschild or EdS) when it reaches $z = 1$ but all light rays are initialized in the EdS region at $\xi = \xi_0$, i.e. the observer is always placed in the EdS region at $\xi = \xi_0$.
\newline\newline
\begin{figure*}
	\centering
	\subfigure[]{
		\includegraphics[scale = 0.45]{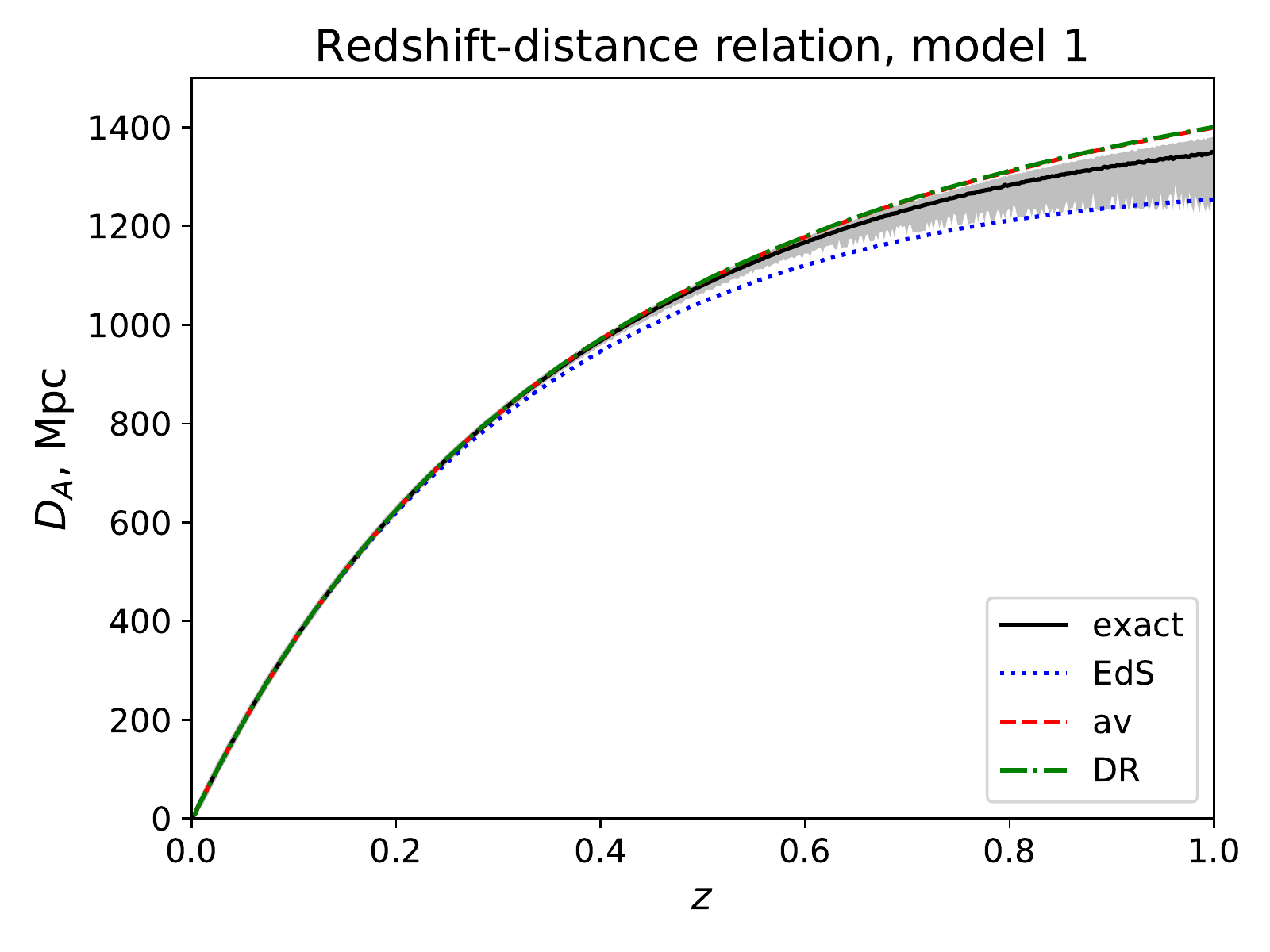}
	}
	\subfigure[]{
		\includegraphics[scale = 0.45]{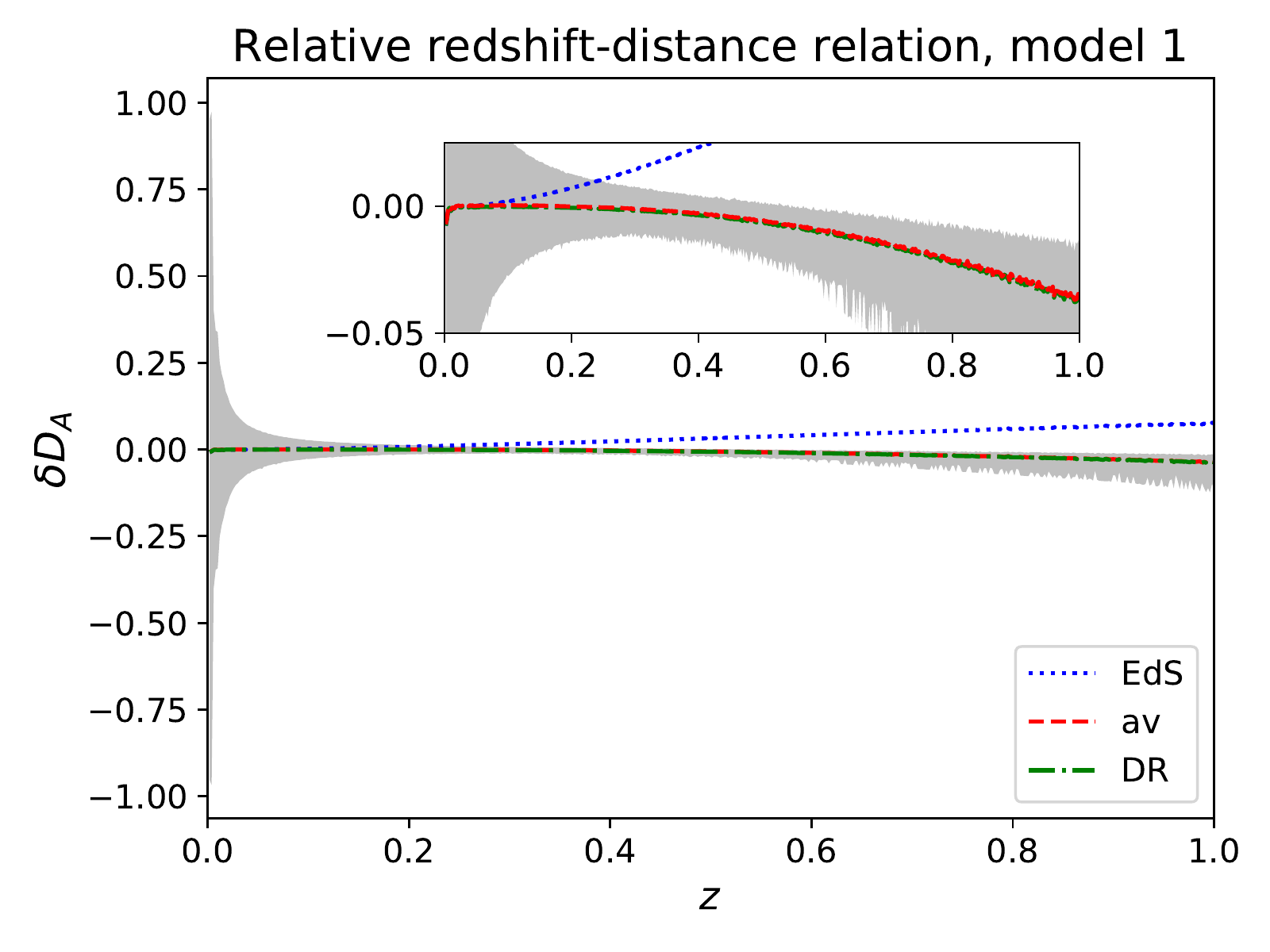}
	}
	\caption{Mean angular diameter distance along 1000 light rays in model 1. The figure to the left shows $D_A$ while the figure to the right shows $\delta D_A:=\frac{D_A-\bar D_A}{\Tstrut\bar D_A}$, where $D_A$ is the exact angular diameter distance along the light rays and $\bar D_A$ is the angular diameter distance according to either the EdS model, the expression based on the spatial averaging scheme (``av'') or the Dyer-Roeder approximation (``DR''). The gray shaded area shows the dispersion around the mean. In the figure to the right, the dispersion is shown for $\frac{D_A-D_{A,DR}}{D_{A,DR}}$.}
	\label{fig:same}
\end{figure*}
Figure \ref{fig:same} shows the mean and dispersion of the redshift-distance relation in model 1 which has $M = 10^{15}M_{\odot}$ and $r_{\rm bie} = 1$Mpc, the inner Schwarzschild region ($r<r_{\rm bie}$) being opaque. The observer was placed in the FLRW region at $\xi_0 = 12.3$Mpc. This situation resembles those studied in \cite{Fleury,Fleuryetal}. As in \cite{Fleury,Fleuryetal}, it is seen that the Dyer-Roeder approximation overestimates the actual mean redshift-distance relation somewhat. As explained in \cite{Fleury}, this is due to shear which is neglected in the Dyer-Roeder approximation and as shown in \cite{Fleury_stochastic} it can partially be corrected for.
\newline\indent
In \cite{Fleury}, it was asserted that spatial averaging is nonsensical in an Einstein-Straus model because there is no natural way to define a 3+1 foliation in the vacuum regions. The parameter $\alpha$ for the Dyer-Roeder approximation was estimated by the fraction of time the light rays spent in the Schwarzschild versus FLRW spacetime regions. Here, it is instead claimed that spatial averaging makes perfect sense in Einstein-Straus models despite there being vacuum regions; the redshift-distance relation of \cite{av_obs1}, reiterated in equations \ref{eq:syksy} and \ref{eq:syksy_z} in the introduction, was specifically derived for averages computed on spatial hypersurfaces of statistical homogeneity and isotropy. It is therefore the distribution of structures that determines the ``natural'' 3+1 foliation for spatial averaging, also in the Einstein-Straus model where there are vacuum regions. Since the Einstein-Straus models studied here are based on distributing Schwarzschild holes randomly in an FLRW cheese, it is the hypersurfaces of constant $T$ which represent the hypersurfaces of statistical homogeneity and isotropy and hence the hypersurfaces on which the spatial averages should be computed.
\newline\indent
Figure \ref{fig:same} serves to explicitly show that, in fact, the Dyer-Roeder approximation and the relation based on spatial averages give the same prediction for the behavior of the mean redshift-distance relation for this particular model - and it also does so in the models studied in \cite{Fleury,Fleuryetal}. To understand why the two procedures give the same prediction, it is necessary to compute the spatially averaged quantities on the spatially spherical region corresponding to $\xi\leq \xi_0$, excluding the opaque region at the center. To do this, first note that the deviation of $A$ from unity is sub-percent for at least $0.1\lesssim r$, so the hypersurfaces of $t = $const. make a good approximation to the hypersurfaces corresponding to $T = $const.. Therefore, the spatial averages on the relevant region ($\xi\leq \xi_0$, $r \geq r_{\rm bie}$) are the spatial averages on the surface with constant $T$ in the region $\xi_{\rm b}\leq \xi\leq \xi_0$ plus the spatial averages on the surface with constant $t$ in the region $r_{\rm bie}\leq r\leq r_{\rm b}$. The deviations of $B$ from 1 is sub-percent as well, so this is also set to 1 when computing spatial averages. Lastly, since sub-percent accuracy is not of interest in this study, the volume of the opaque inner region may be neglected because it makes up such a small fraction of the entire relevant spatial volume.
\newline\newline
The relevant spatially averaged quantities are now straight forward to compute and are
\begin{equation}\label{eq:rho_av}
	\left\langle \rho\right\rangle = \frac{\int_{\xi_{\rm b}}^{\xi_0}\xi^2d\xi\rho_{\rm EdS}}{\frac{1}{3}\xi_0^3} = \frac{\rho_{\rm EdS}\left( \xi_0^3-\xi_{\rm b}^3\right) }{\xi_0^3}:=\alpha\rho_{\rm EdS}
\end{equation}
\begin{equation}\label{eq:_av_theta1}
	\left\langle \theta \right\rangle = \frac{3H_{\rm EdS}\left( \xi_0^3-\xi_{\rm b}^3\right)}{\xi_0^3} + \left\langle \theta_{\rm b}\right\rangle .
\end{equation}
In the computations, it was utilized that the Schwarzschild metric is stationary so has $\theta = 0$, while $\theta = 3H_{\rm EdS}$ in the EdS region.
\newline\indent
Equation \ref{eq:rho_av} gives the prediction for the Dyer-Roeder parameter $\alpha$ used when computing the Dyer-Roeder approximated redshift-distance relation for figure \ref{fig:same}, namely $\alpha =\frac{ \xi_0^3-\xi_{\rm b}^3}{\xi_0^3}$. Equation \ref{eq:_av_theta1} highlights that in order to compute the average expansion rate determining $z_{\rm av}$ (see equation \ref{eq:syksy_z}), the contribution from the boundary expansion rate, $\theta_{\rm b}$, must be obtained. This boundary term appears because the velocity field of comoving observers changes abruptly on the boundary between the Schwarzschild and EdS regions. As shown in appendix \ref{app:Boundary}, such finite jumps can be modeled through a Heaviside function which upon taking a derivative (to go to the expansion rate) becomes a delta function, indicating that the boundary itself must be attributed an expansion rate. The value of the boundary expansion rate can be estimated by considering a Lemaitre-Tolman-Bondi (LTB) model \cite{LTB1,LTB2,LTB3} which for $\xi>\xi_{\rm b}$ expands as an EdS model while being static for $\xi<\xi_{\rm b}$. This estimate is computed in appendix \ref{app:Boundary} where it is found to be
\begin{equation}
\theta_{\rm b} = \xi\delta(\xi-\xi_{\rm b})H_{\rm EdS},
\end{equation}
where $\delta(\xi-\xi_{\rm b})$ is the delta function centered at $\xi = \xi_{\rm b}$.
\newline\newline
Inserting this estimate for $\theta_{\rm b}$ into equation \ref{eq:_av_theta1} yields
\begin{equation}\label{eq:av_theta_small_interior}
\begin{split}
\left\langle\theta \right\rangle = \frac{(\xi_{0}^3-\xi_{b}^3)H_{\rm EdS} + \int_0^{\xi_0}d\xi H_{\rm EdS}\delta(\xi-\xi_{\rm b})\xi^3}{\frac{1}{3}\xi_0^3}\\ =  \frac{(\xi_{0}^3-\xi_{b}^3)H_{\rm EdS} + \xi_{\rm b}^3H_{\rm EdS}}{\frac{1}{3} \xi_0^3} = 3H_{\rm EdS}.
\end{split}
\end{equation}
Thus, the volume averaged expansion rate on surfaces of statistical homogeneity and isotropy is that of the EdS model
\footnote{Note that this result must be insensitive to the exact foliation used in the Schwarzschild region as long as the spatial hypersurfaces only deviate little from the $t = $ const. hypersurfaces. It is possible that hypersurface foliations fulfilling this requirement as well as having a smooth expansion rate across the boundary between the Schwarzschild and FLRW regions exist, whereby the result can be obtained without introducing a boundary contribution, $\theta_{\rm b}$. This e.g. seems to be the case for the hypersurfaces orthogonal to the free-falling velocity fields discussed in \cite{Fleury}.}
. This implies that the naive extension of the analyses of \cite{av_obs1,av_obs2,Linder} to include opaque regions predicts that, to a good approximation and for this particular setup, the mean redshift-distance relation is the same as in the Dyer-Roeder approximation with the equation for $D_A$ modified by $\alpha = \frac{\xi_0^2-\xi_{\rm b}^3 }{\xi_0^3}$ and the redshift given according to the background.
\newline\newline
The results presented above show that the Dyer-Roeder approximation and the competing results based on spatial averages actually agree in the cases studied above and in \cite{Fleury,Fleuryetal}. These setups are therefore ill-suited for distinguishing between the two approaches with explicit examples. In order to distinguish between the two approaches, a model that leaves $\left\langle \theta\right\rangle \not\approx 3H_{\rm EdS}$ must be considered. This is easily achieved by extending the opaque, interior Schwarzschild region. For concreteness, an Einstein-Straus model with $r_{\rm bie}= 10$Mpc, $\xi_0 = 26.1$Mpc and $M = 10^{16}M_{\odot}$ has been studied (model 2). The large mass is introduced in order to exaggerate the effects of the inhomogeneities. The spatial averages must now be computed using the more accurate formulas (still neglecting the deviations of $A,B$ from 1)
\begin{equation}
\left\langle \rho\right\rangle = \rho_{\rm EdS}\frac{\xi_0^3-\xi_{\rm b}^3 }{\xi_0^3-\xi_{\rm bie}^3}:=\alpha \rho_{\rm EdS}
\end{equation}
\begin{equation}
\left\langle\theta \right\rangle =  3H_{\rm EdS}\frac{\xi_{0}^3}{\xi_0^3-\xi_{\rm bie}^3},
\end{equation}
where $\xi_{\rm bie}:=r_{\rm bie}/a$.
\newline\newline
\begin{figure*}
	\centering
	\subfigure[]{
		\includegraphics[scale = 0.45]{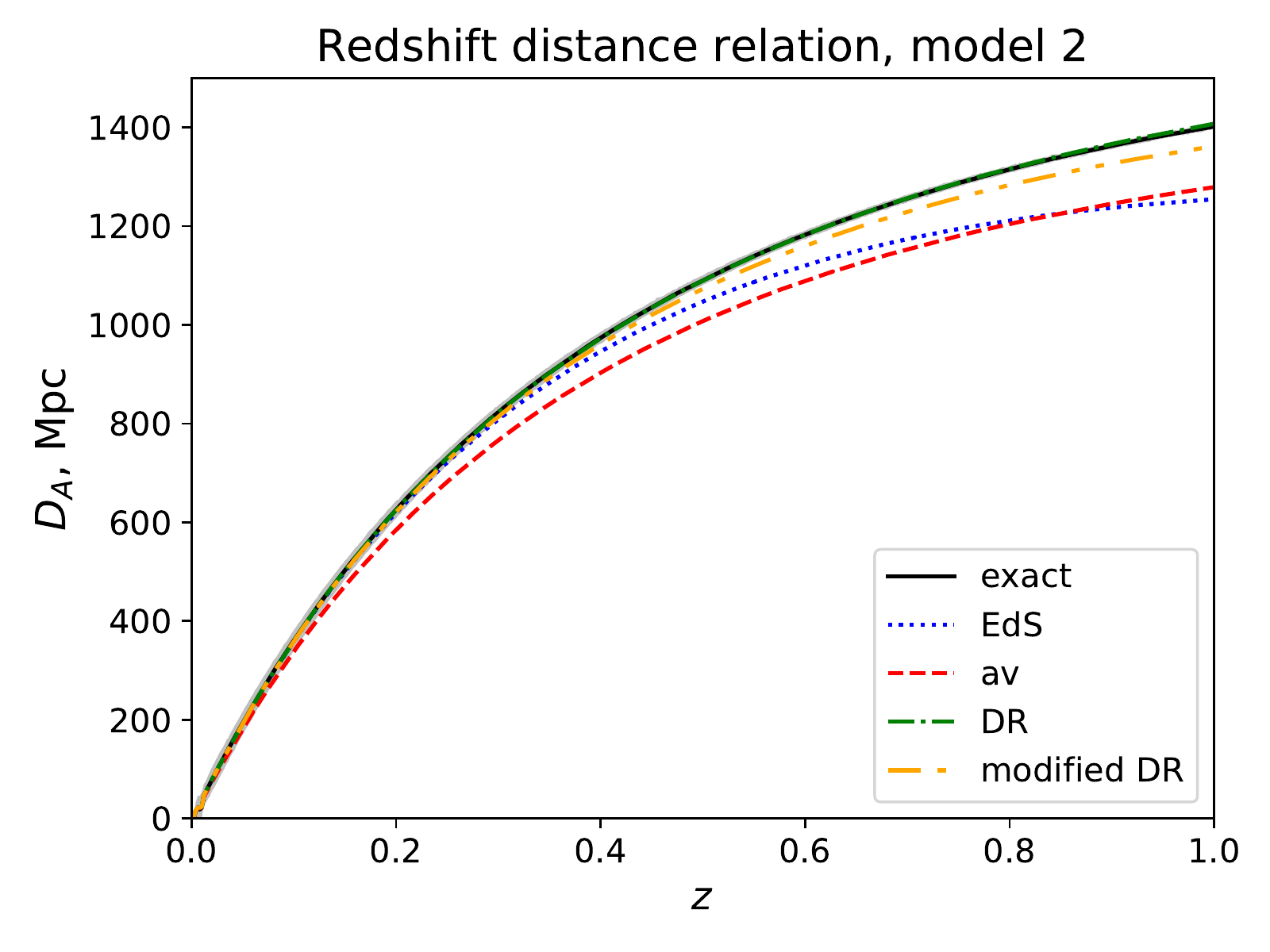}
	}
	\subfigure[]{
		\includegraphics[scale = 0.45]{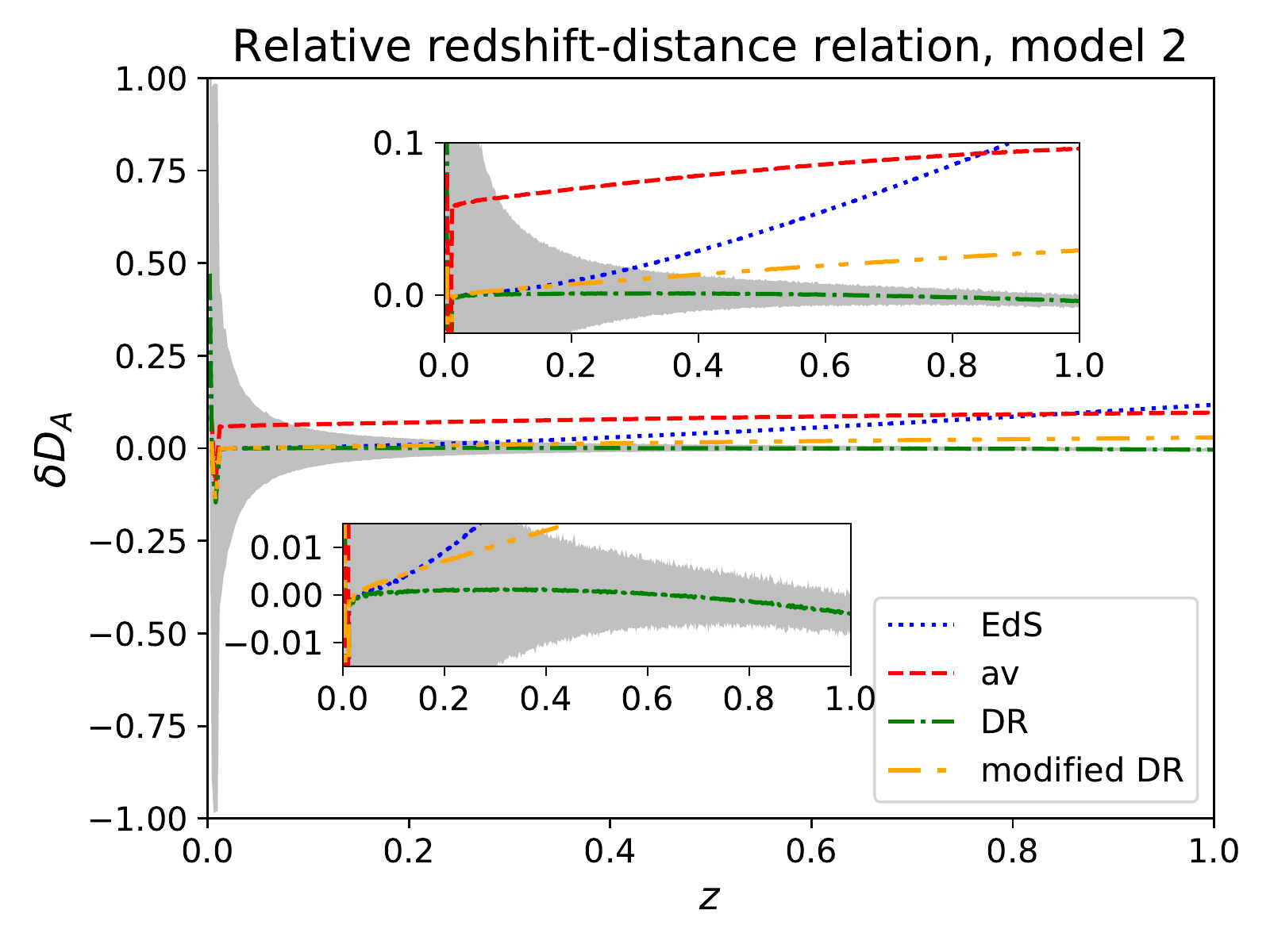}
	}
	\caption{Mean angular diameter distance along 1000 light rays in model 2. The figure to the left shows $D_A$ while the figure to the right shows $\delta D_A:=\frac{D_A-\bar D_A}{\Tstrut\bar D_A}$, where $D_A$ is the exact angular diameter distance along the light rays and $\bar D_A$ is the angular diameter distance according to either the EdS model, the expression based on the spatial averaging scheme (``av''), the Dyer-Roeder approximation (``DR'') or a modified Dyer-Roeder scheme (``modified DR''). A shaded area is included in the figure to the right to show the dispersion around the mean when $\bar D_A$ is given according to the Dyer-Roeder approximation. A dispersion is not included in the figure to the left as it would not be discernible.  Close-ups are shown in the figure to the right.}
	\label{fig:bigOpaque}
\end{figure*}
This time, the average expansion rate is significantly different from the EdS expansion rate and hence the volume averaged scale factor and corresponding redshift will deviate from the EdS redshift. The prediction of the spatially averaged redshift-distance relation therefore differs from that of the Dyer-Roeder approximation. This is seen in figure \ref{fig:bigOpaque} where the mean and dispersion of the redshift-distance relation along 1000 light rays in model 2 are shown. As seen, the Dyer-Roeder approximation makes a good approximation of the mean redshift-distance relation while the predictions based on spatially averaged quantities are clearly wrong. Following critique given of the Dyer-Roeder approximation in e.g. \cite{cc3,misinterp}, a mix between the Dyer-Roeder approximation and the method based on spatial averages, where all quantities in the equation for the angular diameter distance are the spatial averages while the redshift is the EdS redshift, is also included. This `modified Dyer-Roeder approximation' does a poorer job than the ordinary Dyer-Roeder approximation does.
\newline\newline
\begin{table}
	\centering
	\begin{tabular}{c c}
		\hline\hline\Tstrutbig
		Model & $\alpha$\\
		\hline\Tstrutbig
		1& $\sim 0.06$ \\
		2&  $\sim 0.01-0.02$ \\
		3&   $\sim 0.25-0.33$\\
		\hline
	\end{tabular}
	\caption{Value of $\alpha$ in mean-redshift interval $z\in[0,1]$ for the studied models. Values are approximate as indicated by ``$\sim$''.}
	\label{table:alpha}
\end{table}
Note now that in both the models studied so far, $\alpha$ has been quite small. This means that the mean redshift-distance relations in models 1 and 2 are nearly those of empty beams. It is instructive to also look at a model which has a somewhat larger $\alpha$. For this purpose, model 3 is constructed using the same parameters as model 2 except that it includes a larger FLRW volume. $\alpha$ for all three models is shown in table \ref{table:alpha}. Note that $\alpha$ varies in time because $\xi_{\rm bie}$ does.
\newline\indent
Figure \ref{fig:bigalpha} shows the redshift-distance relation for model 3. The Dyer-Roeder approximation clearly still gives a much better approximation of the actual mean redshift-distance relation than the relations based on spatial averages and is also still better than the modified Dyer-Roeder approximation (though only barely). However, the difference between the Dyer-Roeder approximation and the actual mean redshift-distance relation is more significant than it was for models 1 and 2. This indicates that the spatial average used for computing $\alpha$ does not exactly capture the spacetime felt by the light rays on average; any such imprecision in averages will for the Dyer-Roeder approximation only affect $\alpha$ and hence does not have much impact when $\alpha$ is close to zero but will become visible when $\alpha$ becomes of order $0.1-1$.
\newline\indent
It is not surprising that the spatial averages do not coincide {\em exactly} with the ``mean'' spacetime felt by the light rays. First of all, the spatial averages on hypersurfaces of statistical homogeneity and isotropy can only in the case of a stationary metric be expected to correspond exactly to the mean felt by light rays. Second of all, as already discussed, the hypersurfaces used here for computing spatial averages and hence $\alpha$ are only approximately those of spatial homogeneity and isotropy. Lastly, while it is only the spherical regions for which $r\leq r_{\rm bie}$ which are modeled as opaque, these opaque spheres effectively block an entire cone of the spacetime $\xi\leq\xi_0$ which a light ray cannot traverse. The correct volume averages must therefore be modified compared to those used above where only the spherical regions were considered opaque. However, the good agreement between the Dyer-Roeder approximation and mean redshift-distance relation in model 3 indicates that the spatial averages used here constitute quite good approximations of the average spacetime probed by the light rays.
\newline\newline
Since the redshift is the quantity which most clearly differs between the Dyer-Roeder approximation and the relations based on spatial averages, it is instructive to look at the redshift in more detail. This in done in the following subsection.

\begin{figure*}
	\centering
	\subfigure[]{
		\includegraphics[scale = 0.45]{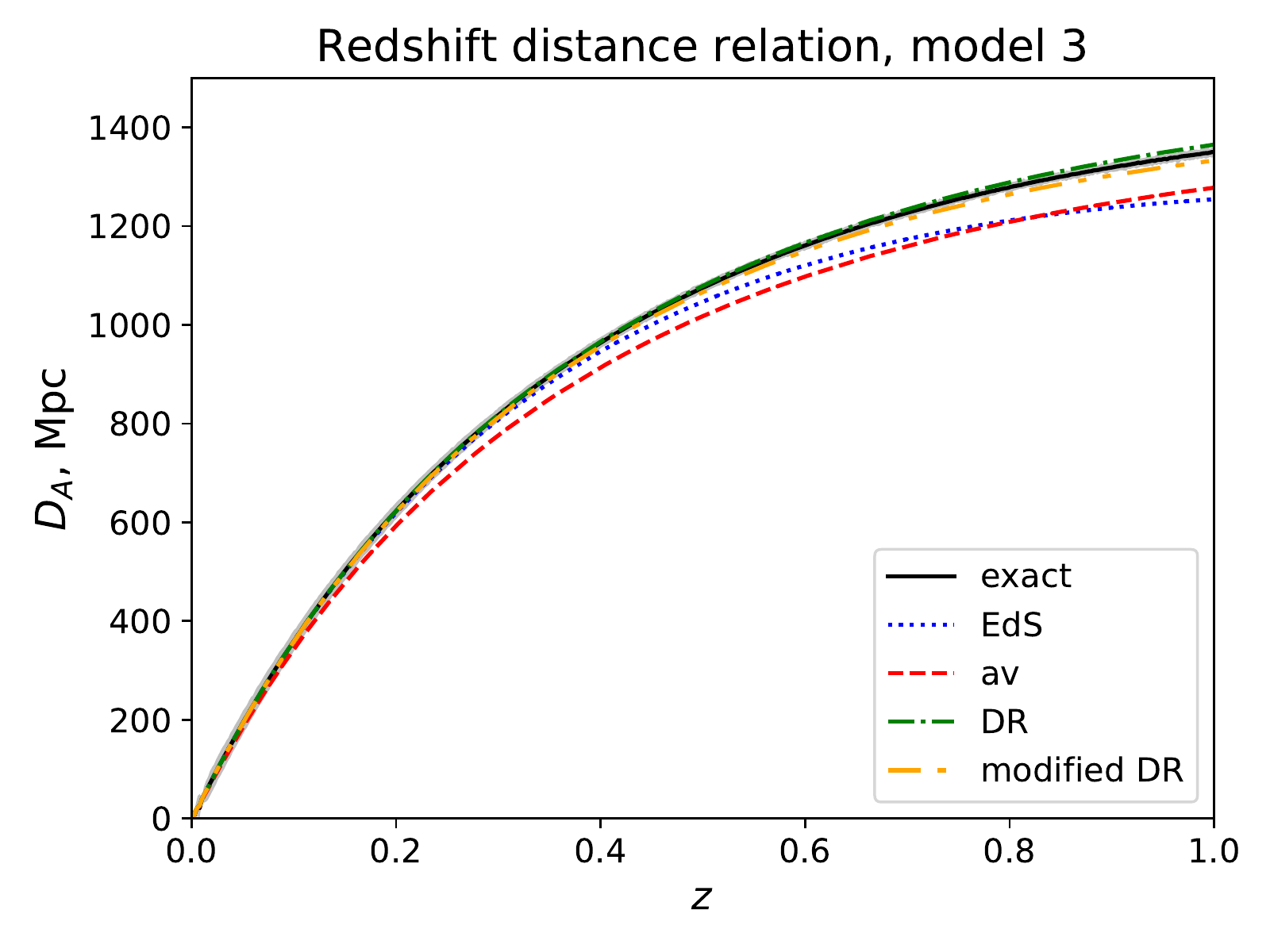}
	}
	\subfigure[]{
		\includegraphics[scale = 0.45]{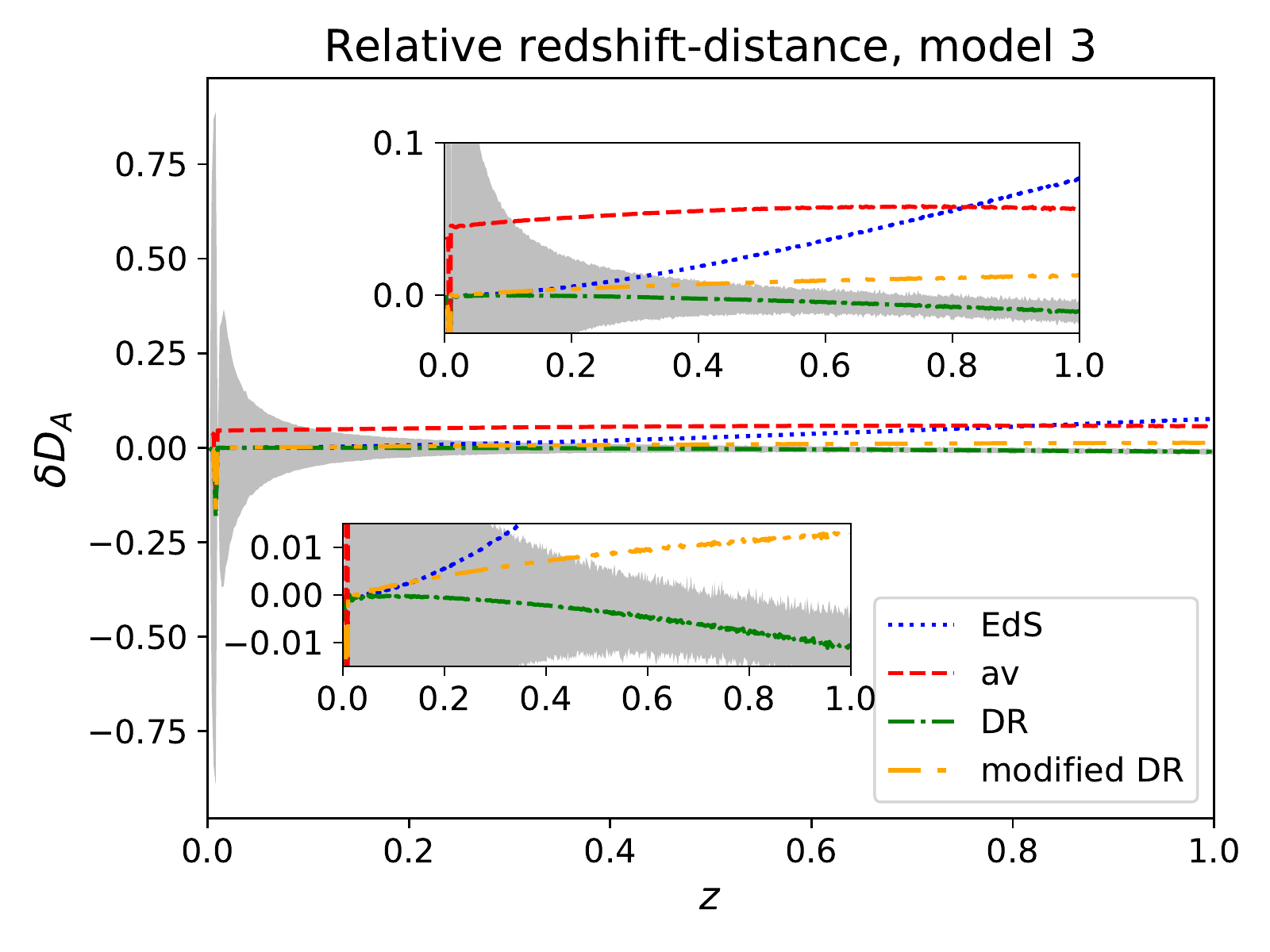}
	}
	\caption{Mean angular diameter distance along 1000 light rays in model 3. The figure to the left shows $D_A$ while the figure to the right shows $\delta D_A:=\frac{D_A-\bar D_A}{\Tstrut\bar D_A}$, where $D_A$ is the exact angular diameter distance along the light rays and $\bar D_A$ is the angular diameter distance according to either the EdS model, the expression based on the spatial averaging scheme (``av''), the Dyer-Roeder approximation (``DR'') or a modified Dyer-Roeder scheme (``modified DR''). A shaded area is included in the figure to the right to show the dispersion around the mean when $\bar D_A$ is given according to the Dyer-Roeder approximation. A dispersion is not included in the figure to the left as it would not be discernible.  Close-ups are shown in the figure to the right.}
	\label{fig:bigalpha}
\end{figure*}

\subsection{Redshift contributions}
\begin{figure}
	\centering
		\subfigure[]{
	\includegraphics[scale = 0.45]{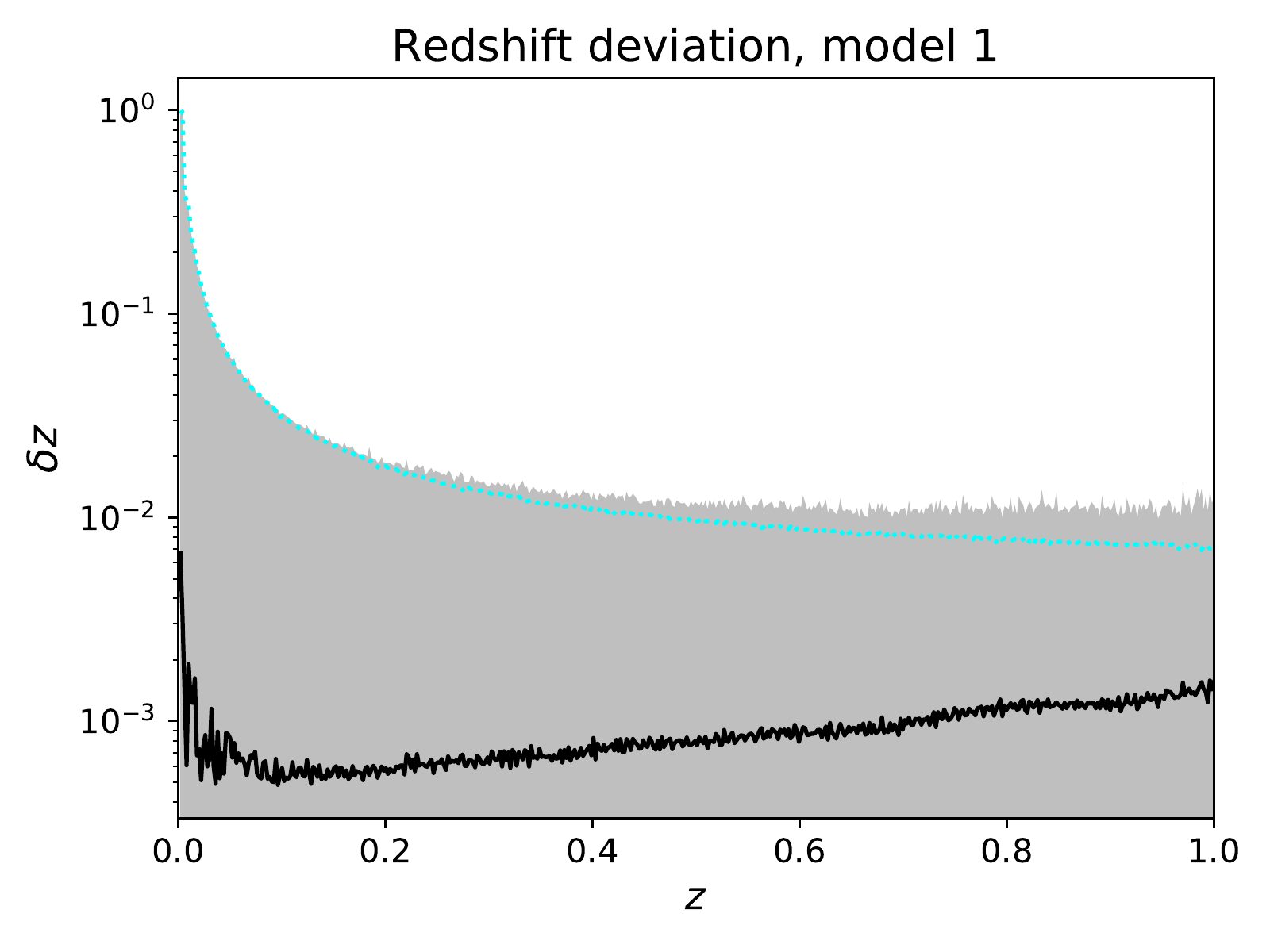}
	}
	\subfigure[]{
		\includegraphics[scale = 0.45]{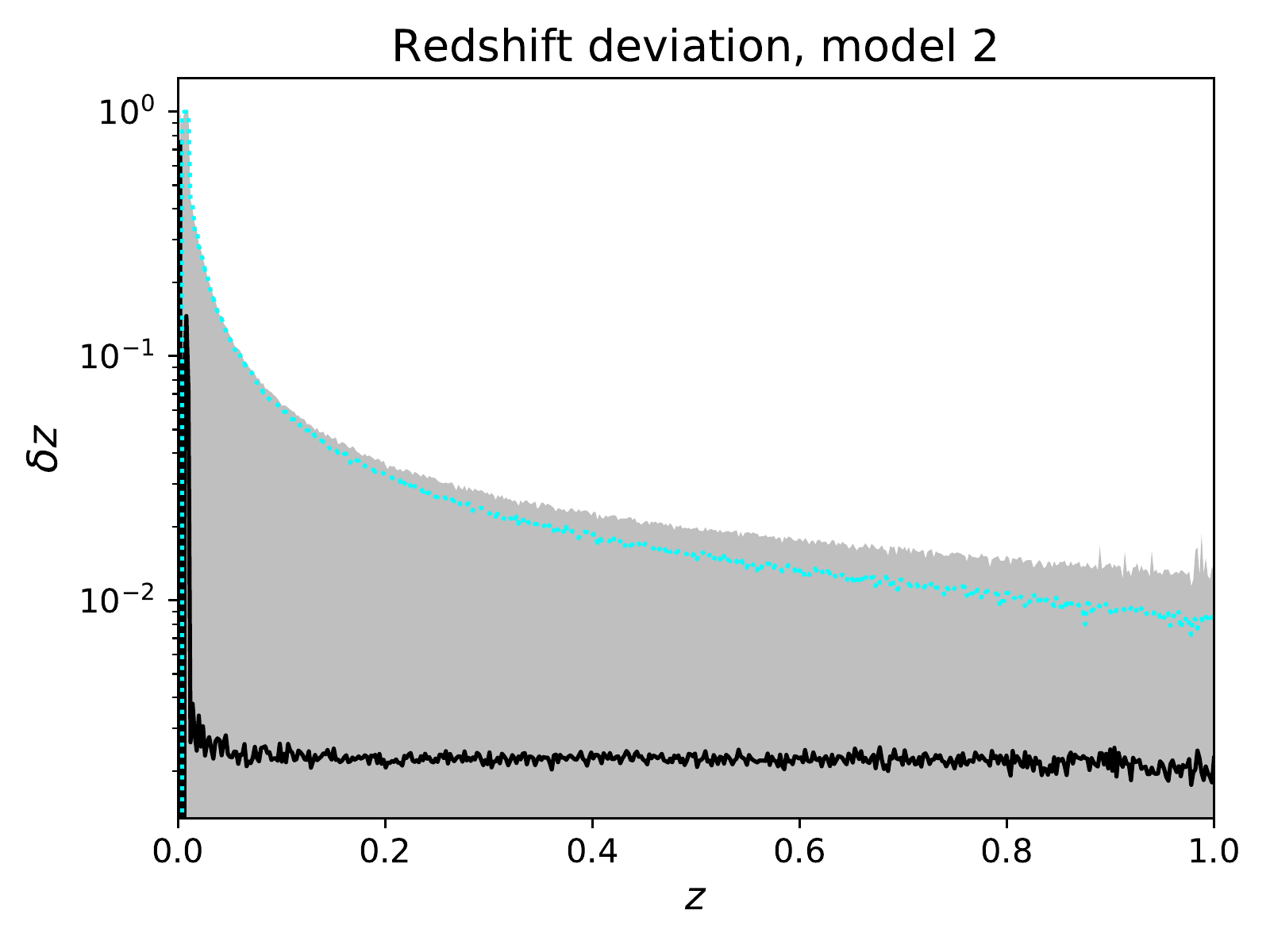}
	}
	\subfigure[]{
	\includegraphics[scale = 0.45]{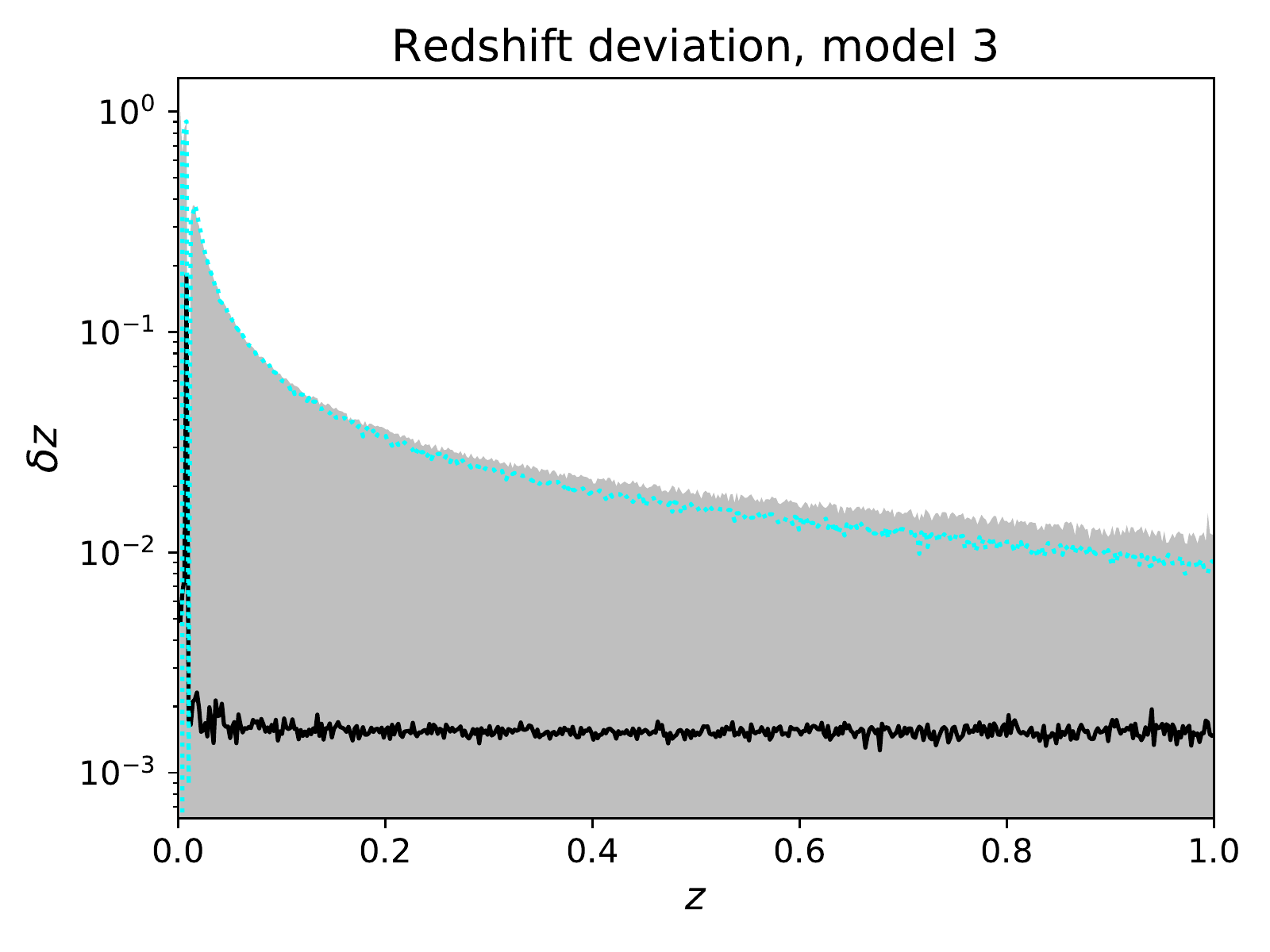}
	}
	\caption{Fluctuations in the redshift, $\delta z:=\frac{z_{\rm exact}-z_{\rm EdS}}{z_{\rm exact}}$, along 1000 light rays for models 1, 2 and 3. The black line shows the mean while the shaded area shows the  dispersion. The negative valued lower rim of the dispersion is indicated in cyan through absolute values.}
	\label{fig:dz_bigOpaque}
\end{figure}

\begin{figure*}
	\centering
	\subfigure[]{
		\includegraphics[scale = 0.45]{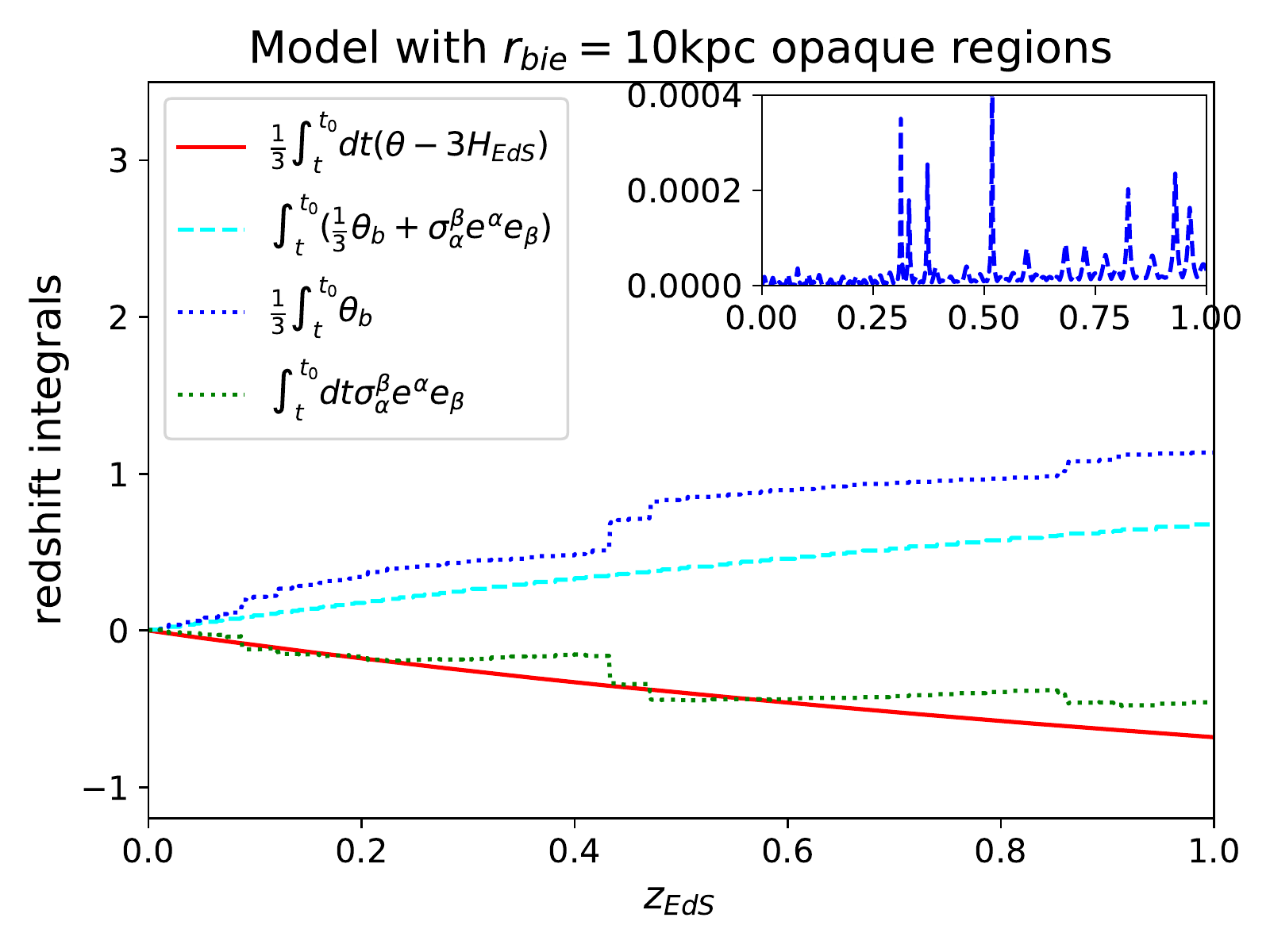}
	}
	\subfigure[]{
		\includegraphics[scale = 0.45]{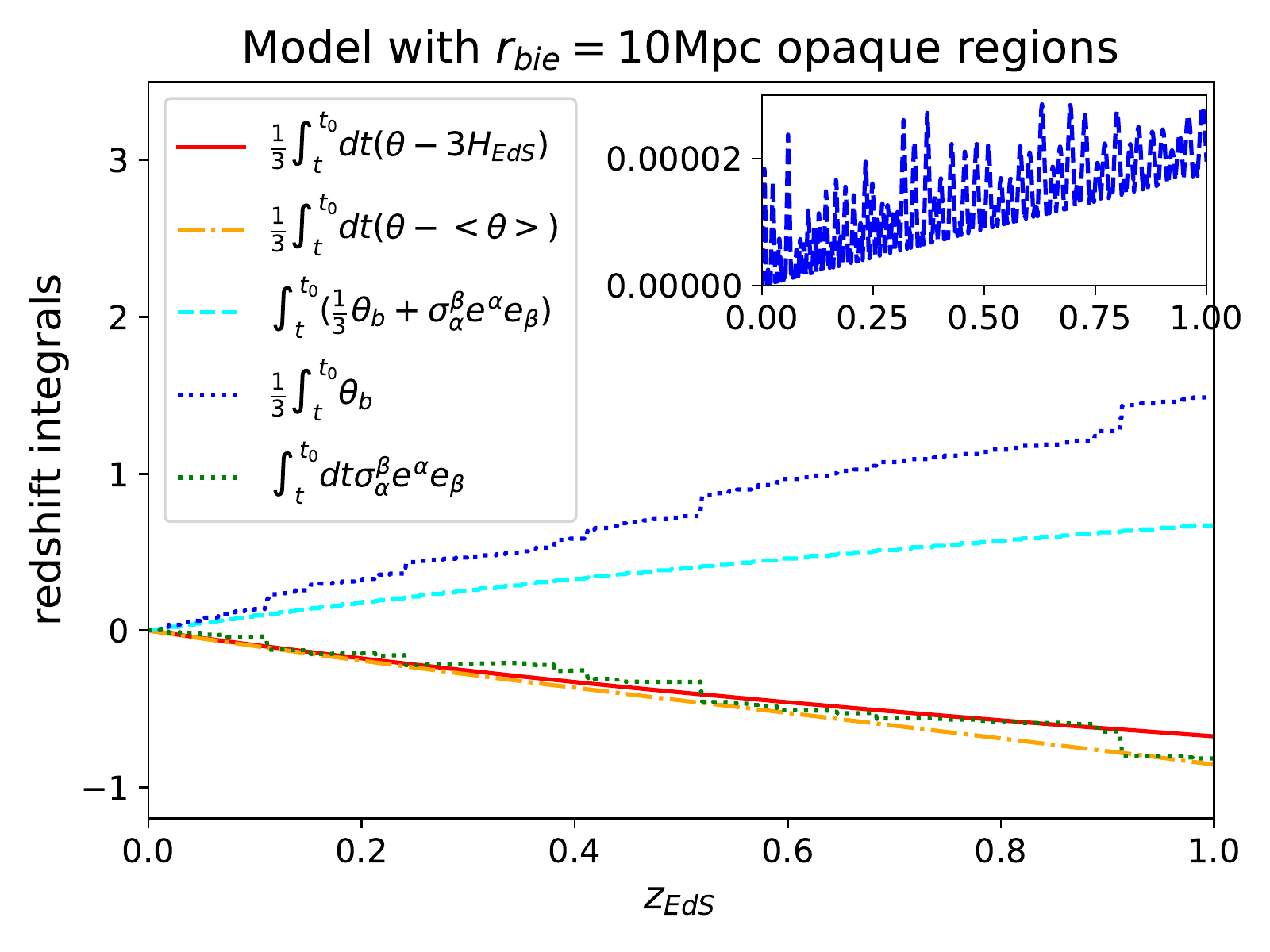}
	}
	\subfigure[]{
		\includegraphics[scale = 0.45]{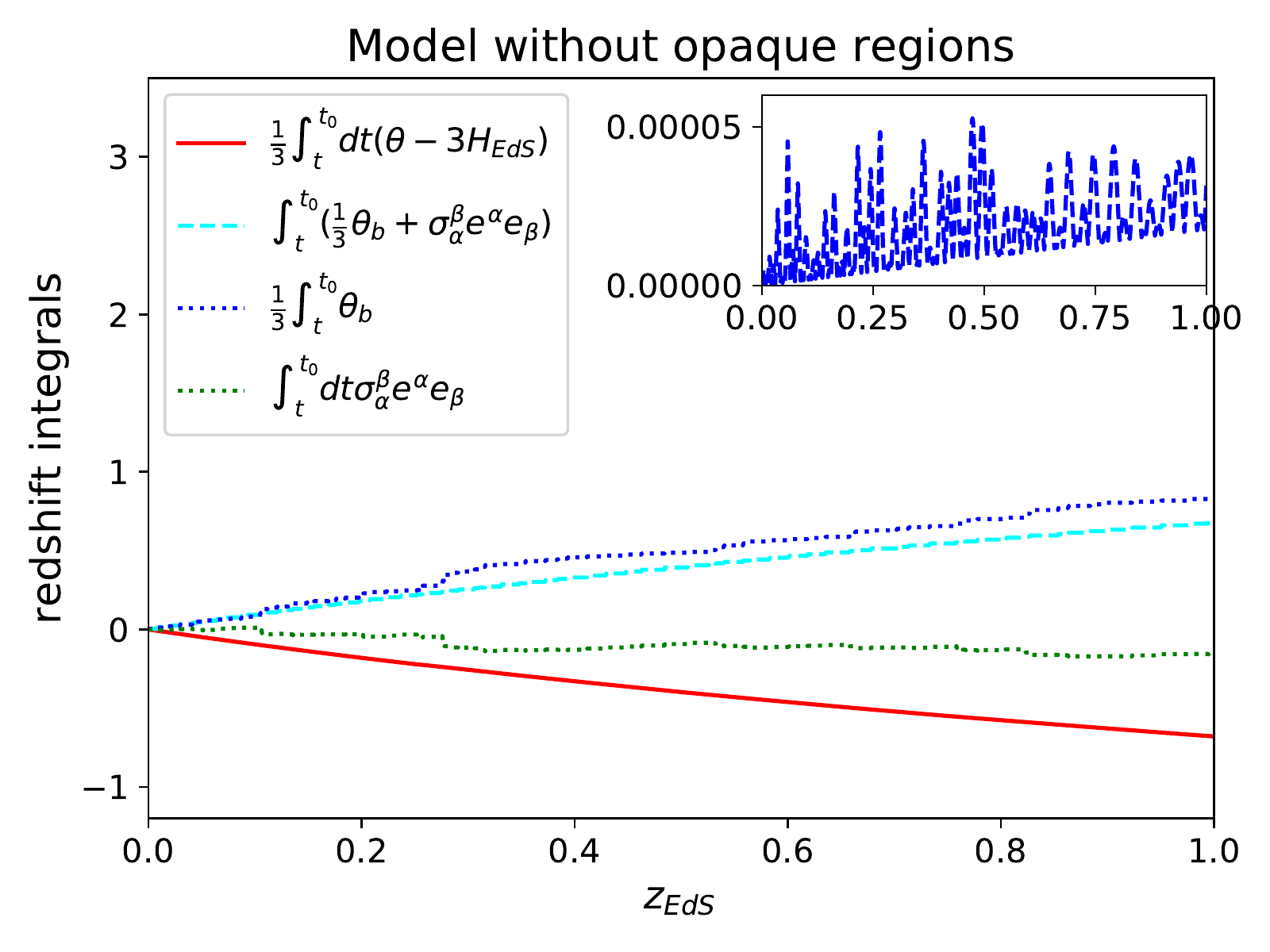}
	}
	\caption{Redshift components along individual light rays for models with $M = 10^{16}M_{\odot}$, $\xi_0 = 26.1$Mpc and different values of $r_{\rm bie}$. The contribution from the acceleration, $1+z_{\rm Sch}$, is nearly zero so is not included in the main figures but is instead shown in separate close-ups in each subfigure. In the top right figure, the contributions from the fluctuations in the shear are shown both with respect to the background value (3$H_{\rm EdS}$) and the spatial average with spatial averages not including opaque regions ($<\theta>$).}
	\label{fig:z_components}
\end{figure*}

Figure \ref{fig:dz_bigOpaque} shows the mean of the deviation in the redshift compared to the EdS redshift for models 1, 2 and 3. As seen, in all three cases, the deviations from the EdS redshift are only around 0.2 \% for $z\approx 1$, so the mean redshift is in all the studied cases very well approximated by the background redshift, $z_{\rm EdS}$. One may however note that there seems to be an offset in the mean redshifts compared to the EdS redshift. This offset is clearest in models 2 and 3. It has been verified that this result is not a statistical fluke by computing the redshift along several different sets of 1000 random light rays. The offset seems to be robust. The origin of this offset has not been identified. As the offset is quite small, it is not of particular interest in this study and will not be considered further.
\newline\newline
In order to better understand why the Dyer-Roeder approximation makes a much better prediction of the redshift-distance relation than the method based on spatial averages does, it is instructive to look at the different contributions to the redshift according to the general expression
\begin{equation}
1+z = \exp\left( \int_t^{t_0}\Gamma dt\left(\frac{1}{3}\theta + \sigma_{\alpha\beta}e^{\alpha}e^{\beta} + \dot u^{\alpha}e_{\alpha} \right)\right) , 
\end{equation}
where $e^{\alpha} = -\frac{k^{\alpha}}{k^{\beta}u_{\beta}}-u^{\alpha}$ is the spatial direction vector of the light ray, $\dot u^{\alpha}:=u^{\beta}\nabla_{\beta}u^{\alpha}$ is the fluid acceleration and $\Gamma$ is the lapse function which is equal to 1 in the EdS region and to $A$ in the Schwarzschild region. These different contributions to the redshift are discussed in detail below and are shown in figure \ref{fig:z_components} along single, random light rays in model 2 and variations of model 2 with different values of $r_{\rm bie}$.
\newline\indent

\begin{figure*}
	\centering
	\subfigure[]{
		\includegraphics[scale = 0.45]{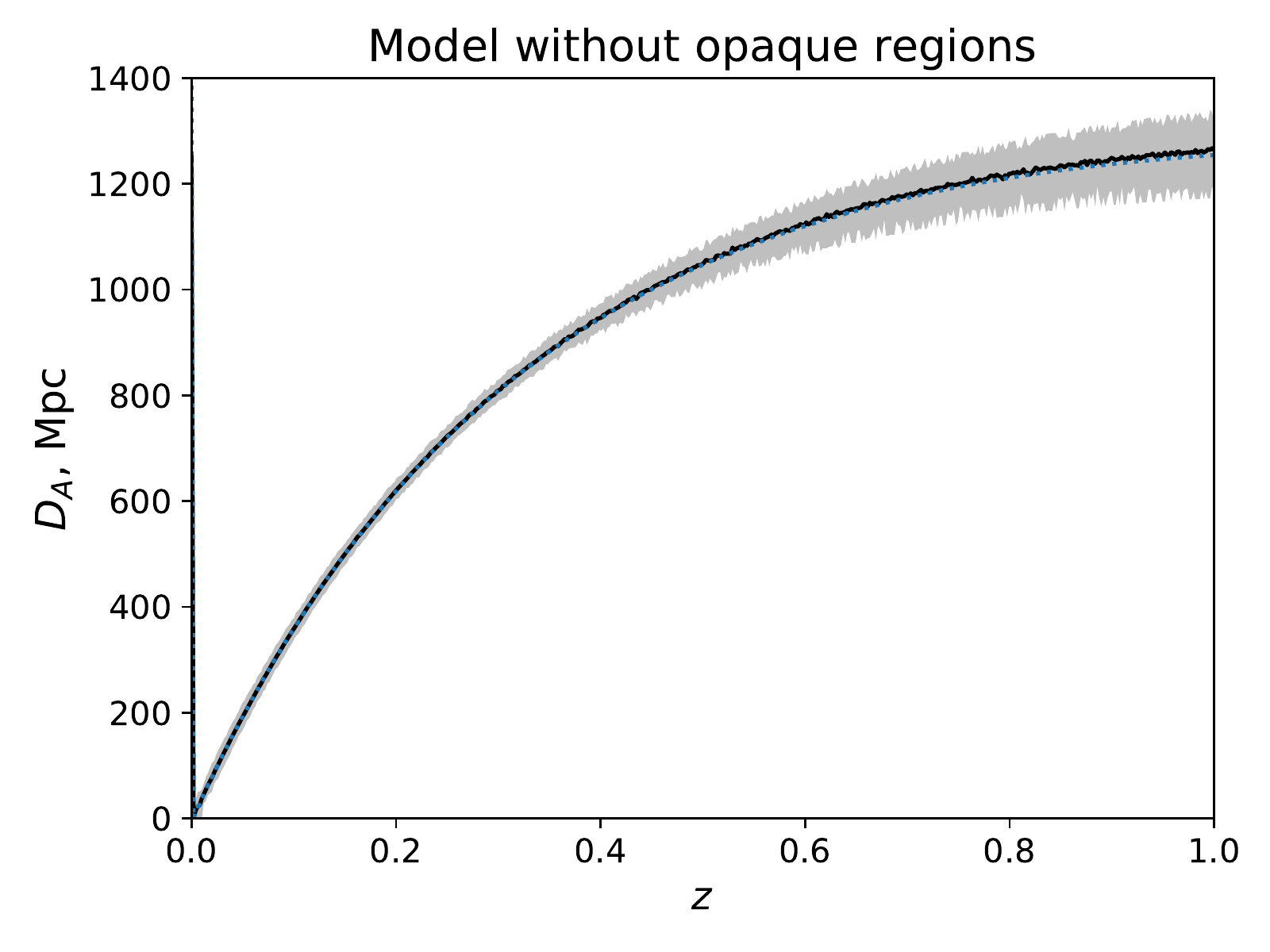}
	}
	\subfigure[]{
		\includegraphics[scale = 0.45]{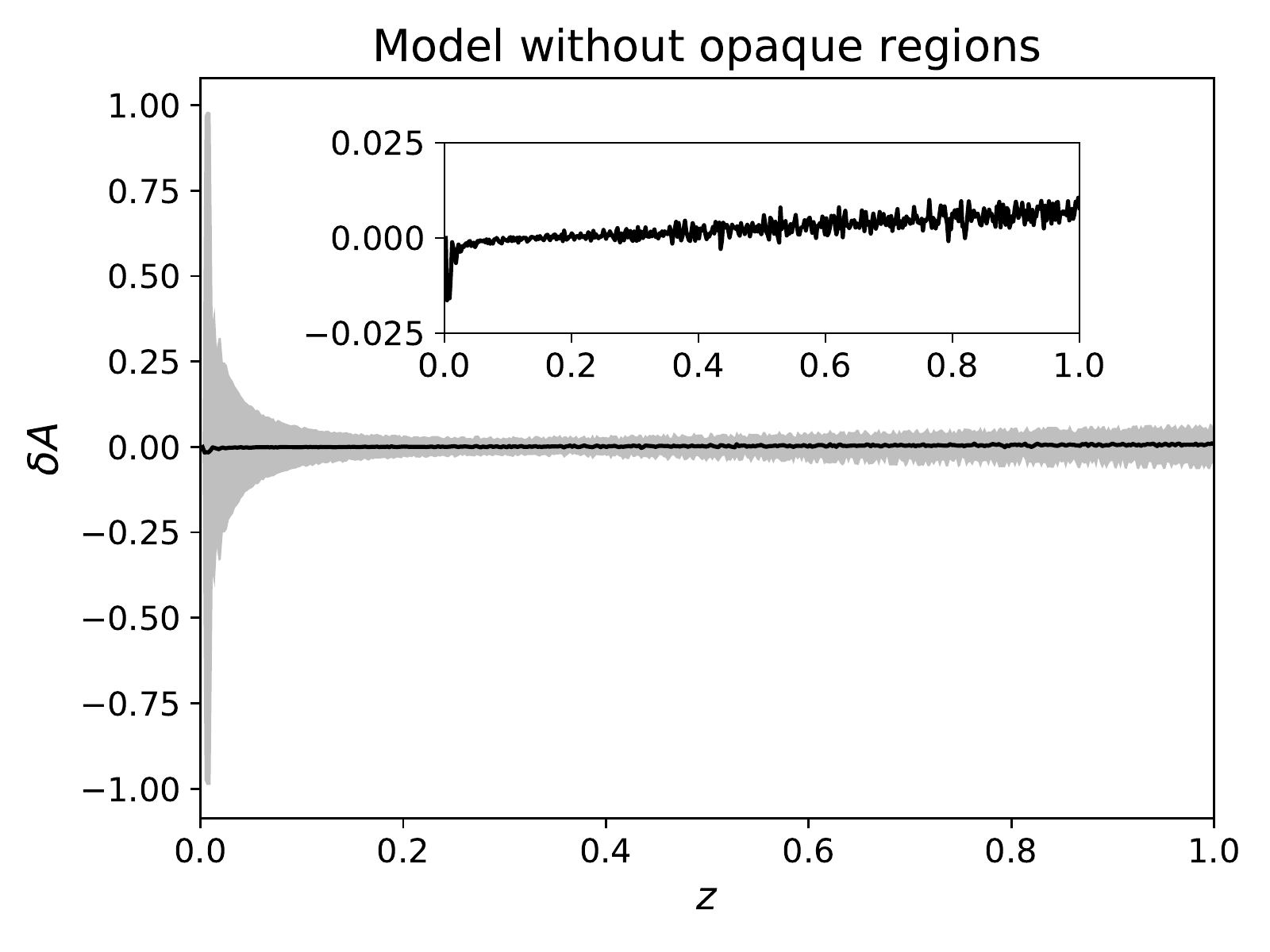}
	}
	\caption{Mean angular diameter distance along 1000 light rays in Einstein-Straus model with transparent interior regions of $M = 10^{16}M_{\odot}$ and $r_{\rm bie} = 10$Mpc (i.e. model 2 with interior regions made transparent). The figure to the left shows $D_A$ while the figure to the right shows $\delta D_A:=\frac{D_A- D_{A,\rm EdS}}{D_{A,\rm EdS}}$. A shaded area is included to show the dispersion around the mean. A close-up is also shown in the figure to the right. The figure to the left shows $D_A$ as a black line and $D_{A,\rm EdS}$ as a cyan dotted line but the two lines overlap so precisely that they look like a single (black) line.}
	\label{fig:DA_shear}
\end{figure*}

In the Schwarzschild region, the only contribution comes from the acceleration of the source. In the exterior Schwarzschild region there is no natural velocity field to choose but in the interior region, the natural choice would be the velocity of the (comoving) fluid. This velocity field is $u_{\mu}^{\rm sch} = \left(-\sqrt{A},0,0,0\right) $ with corresponding acceleration $\dot u_{\mu}^{\rm Sch} = \left(0,\frac{1}{2}\frac{A_{,r}}{A},0,0 \right) $. This velocity field is also chosen in the exterior Schwarzschild metric. In \cite{Fleury,Fleuryetal}, another velocity field was chosen, namely one of a free-falling observer. It does not matter much which velocity field is chosen in the Schwarzschild region; the difference between the redshift seen by different observers is a matter of a local Doppler term so the choice will not affect the redshift of light emitted by a comoving source in the EdS region and seen by the comoving observer also in the EdS region and it therefore does not affect the mean redshift significantly. One may note though that for a non-vacuum region, the velocity field of the fluid component is natural to choose not only because it is the velocity field representing the only actual source in the model, but also because this source directly affects the spacetime geometry and dynamics and therefore its velocity can be expressed through the metric functions.
\newline\indent
The redshift contribution from a single traversal of a Schwarzschild region is given by
\begin{align}
\begin{split}
	1+z_{\rm sch} &= \exp\left( \int_{t_i}^{t_o}dt\sqrt{A}\dot u^{\alpha}e_{\alpha}\right)  = \exp\left( -\frac{1}{2}\int_{t_i}^{t_o}dt\sqrt{A}\frac{A_{,r}}{A}\frac{k^r}{u_t^{\rm Sch}k^t}\right)  \\ &= 
			\exp\left( \int_{r_i}^{r_o}dr\frac{1}{2}\frac{A_{,r}}{\sqrt{A}}\right)  = 	\exp\left( \int_{r_i}^{r_o}dr\frac{d}{dr}\ln\sqrt{A} \right)  =\sqrt{\frac{A(r_{\rm b}(t_o))}{A(r_{\rm b}(t_i))}} = \sqrt{\frac{1-\frac{2M}{r_{\rm b}(t_o)}}{1-\frac{2M}{r_{\rm b}(t_i)}}},
\end{split}
\end{align}
where the subscript $i$  indicates entry into the Schwarzschild region and $o$ indicates exit out of it. The second equality is based on noting that the spatial part of $e^{\mu}$ is simply $e^{i} = \frac{k^i}{-k^t\sqrt A}$.
\newline\indent
The expression for $1+z_{\rm Sch}$ found above is just the ordinary expression for the redshift in the Schwarzschild metric as seen by a comoving observer (observing a comoving source). The overall contribution to the redshift upon traversing an entire Schwarzschild region is purely due to the expansion of the Schwarzschild region's boundary and is hence an integrated Sachs-Wolfe/Rees-Sciama effect \cite{ISW,ReesSciama, ISWReesSciama_review} as was also noted in \cite{Fleury}. Note that the contribution fluctuates as the light ray traverses a Schwarzschild hole and that it is only upon traversal of an {\em entire} hole that the contribution can be considered an ISW/Reese-Sciama effect.
\newline\newline 
In the EdS region, the only contribution to the redshift is that due to the expansion. Thus, the contribution to the redshift from traversing one EdS region is
\begin{equation}
	1+z_{\rm EdS} = \exp\left( \int_{T_1}^{T_2}H_{\rm EdS}dT\right)  = \frac{a_{\rm EdS}(T_1)}{a_{\rm EdS}(T_2)} .
\end{equation}
$T_1$ and $T_2$ denote the time coordinates when the light ray enters and exits the EdS region, respectively.
\newline\newline
The remaining contribution to the redshift comes from the boundary between the EdS and exterior Schwarzschild regions. As with the volume averages, the contribution from the boundary can be estimated by modeling the boundary as an LTB model. The details are given in appendix \ref{app:Boundary} where it is shown that there is a delta function contribution to the shear in addition to the delta function contribution to the expansion rate that was used earlier for computing the volume averaged expansion rate. Specifically, the shear contribution from the boundary is
\begin{equation}
\begin{split}
\sigma_{\alpha}^{\beta} = \xi H_{\rm EdS}\delta(\xi -\xi_{\rm b})\cdot\text{\rm diag}\left(0,\frac{2}{3},-\frac{1}{3},-\frac{1}{3} \right).
\end{split}
\end{equation}
The two contributions to the redshift from crossing the boundary once are thus
\begin{align}\label{eq:z_boundary1}
\begin{split}
	1+z_{\theta_{\rm b}} &= \exp\left( \frac{1}{3}\int_{T_1}^{T_2}dT\xi H_{\rm EdS}\delta(\xi - \xi_{\rm b})\right)  \\ &=
	\exp\left( \frac{1}{3}\int_{\xi_1}^{\xi_2}d\xi \frac{k^T}{k^{\xi}}\xi H_{\rm EdS}\delta(\xi-\xi_{\rm b})\right)  \\&= \exp\left( \frac{1}{3}\left| \frac{k^T}{k^{\xi}}\right|_{\rm b} \xi_{\rm b}H_{\rm EdS}\right) 
	\end{split}
\end{align}
and
\begin{align}
\begin{split}
	1+z_{\sigma_{\rm b}} &= \exp\left( \int_{T_1}^{T_2}dT\sigma_{\alpha}^{\beta}e^{\alpha}e_{\beta}\right)  \\ & =
	\exp\left( \int_{T_1}^{T_2}dT\left( \frac{a^2H_{\rm EdS}\xi\delta(\xi - \xi_{\rm b})}{3\left( k^T\right) ^2}\left[ 2\left( k^{\xi}\right)^2- \xi^2\left( k^{\theta}\right) ^2 - \xi^2\sin^2(\theta)\left( k^{\phi}\right) ^2\right] \right) \right)   \\ &=
		\exp\left( \int_{\xi_1}^{\xi_2}d\xi\left( \frac{a^2H_{\rm EdS}\xi\delta(\xi-\xi_{\rm b})}{3 \left| k^{\xi}k^T\right| }\left[ 2\left( k^{\xi}\right)^2- \xi^2\left( k^{\theta}\right) ^2 - \xi^2\sin^2(\theta)\left( k^{\phi}\right) ^2\right] \right) \right)  \\&=
	 \exp\left( \frac{a^2H_{\rm EdS}\xi_{\rm b}}{3\left| k^Tk^{\xi}\right| }\left[ 2\left( k^{\xi}\right)^2- \xi_{\rm b}^2\left( k^{\theta}\right) ^2 - \xi_{\rm b}^2\sin^2(\theta)\left( k^{\phi}\right) ^2\right]|_{\rm b} \right) ,
	\end{split}
\end{align}
where the integration limits are arbitrary, though only including the boundary once. The two contributions can be multiplied to form a simple expression by noting that the spherical symmetry of the situation implies that we can without loss of generality consider a light ray traveling in a plane with one of the angles kept constant. Keeping the polar angle constant and equal to $\pi/2$, the null condition then leads to\footnote{The possibility of writing $k^{\alpha}$ this way and using it to write the simple expression for $(1+z_{\sigma_{\rm b}})(1+z_{\theta_{\rm b}})$ given in equation \ref{eq:z_b_Fleury} even for non-radial light rays was pointed out to the author by Pierre Fleury.}
\begin{equation}
k^{\xi} = \pm\frac{1}{a_{\rm EdS}}k^T\sin(\phi_{\rm b})
\end{equation}
\begin{equation}
k^{\phi} = \pm \frac{1}{a_{\rm EdS}\xi_{\rm b}}k^T\cos(\phi_{\rm b}),
\end{equation}
where $\phi_{\rm b}$ is the angle of impact between the light ray and the boundary with $\phi_{\rm b} = \pi/2$ corresponding to a radial light ray. We can therefore write the total boundary contribution to the redshift as
\begin{align}\label{eq:z_b_Fleury}
\begin{split}
	1+z_{\rm b}:&=(1+z_{\sigma_{\rm b}})(1+z_{\theta_{\rm b}}) \\ &=  \exp\left( \frac{1}{3}\left| \frac{k^T}{k^{\xi}}\right|_{\rm b} \xi_{\rm b}H_{\rm EdS}+\frac{H_{\rm EdS}\xi_{\rm b}}{3\left| k^Tk^{\xi}\right| }\left[ 2\left( k^T\sin(\phi_{\rm b})\right)^2- \left( k^T\cos(\phi_{\rm b})\right) ^2\right]|_{\rm b}\right) 
	\\ &=
	\exp\left(a_{\rm EdS}H_{\rm EdS}\xi_{\rm b}\sin(\phi_{\rm b}) \right) .
\end{split}
\end{align}
For each hole a light ray traverses, it crosses the boundary twice, so the total boundary contribution from traversing a single Schwarzschild region is
\begin{equation}
	1+z_{\rm b_2}:=	\exp\left[a_{\rm EdS}(T_{\rm in})H_{\rm EdS}(T_{\rm in})\xi_{\rm b_{in}}\sin(\phi_{\rm b_{in}})+a_{\rm EdS}(T_{\rm out})H_{\rm vEdS}(t_{\rm out})\xi_{\rm b_{\rm out}}\sin(\phi_{\rm b_{out}}) \right],
\end{equation}
where $T_{\rm in}$ and $T_{\rm out}$ are the FLRW time coordinate values at which the light ray enters and exits the Schwarzschild region, respectively. The spherical symmetry means that $\sin(\phi_{\rm b_{in}}) = \sin(\phi_{\rm b_{out}})$ and since the holes are modest in size, $\xi_{b_{in}}\approx \xi_{b_{out}}$. Subscripts $\rm in(out)$ will therefore be omitted on these quantities in the following.
\newline\indent
Note now that, for a light ray in EdS spacetime,
\begin{equation}\label{eq:time_vs_xi}
a_{\rm EdS}(T_{\rm in(out)})\xi_{\rm b}\sin(\phi_{\rm b}) = a_{\rm EdS}(T_{\rm in(out)})\int_{\Delta T_{\rm in(out)}}\frac{dT}{a_{\rm EdS}} \approx \Delta T_{\rm in(out)},
\end{equation}
where $\Delta T_{\rm in(out)}$ is the time it takes for the light ray to travel the comoving distance $\xi_{\rm b}\sin(\phi_{\rm b})$ at $T\approx T_{\rm in(out)}$ in the EdS spacetime. The approximation indicated by ``$\approx $'' in equation \ref{eq:time_vs_xi} corresponds to neglecting the evolution of $a_{\rm EdS}$ during this travel time. The EdS redshift corresponding to the time interval the light ray travels in a Schwarzschild hole can be approximated as
\begin{align}\label{eq:z_edS}
\begin{split}
	1+z_{\rm EdS} &= \exp\left(\int_{T_{\rm in}}^{T_{\rm out}} dTH_{\rm EdS} \right) =\exp\left(\int_{\Delta \xi} d\xi a_{\rm EdS} H_{\rm EdS} \right)\\ &\approx \exp\left( \xi_{\rm b}\sin(\phi_{\rm b})\left[a_{\rm EdS}(T_{\rm in}) H_{\rm EdS}(T_{\rm in}) + a_{\rm EdS}(T_{\rm out})H_{\rm EdS}(T_{\rm out})\right]  \right) ,
	\end{split}
\end{align}
where $\Delta \xi$ represents the change in the coordinate $\xi$ along the light ray. The approximation indicated by ``$\approx$'' corresponds to assuming (again) that $a_{\rm EdS}$ and $H_{\rm EdS}$ do not change during travel time into and out of the hole. In addition, the approximation requires that the travel time into and out of the Schwarzschild hole actually corresponds to $\Delta T_{\rm in(out)}$ of equation \ref{eq:time_vs_xi}. The latter approximation becomes better the further a light ray is from the central structure. The former approximation becomes better the less time a light ray spends in a given hole. In both cases, the approximation thus becomes better the further the light ray is from being radial. The small values of $\delta z$ in figure \ref{fig:dz_bigOpaque} indicates that the approximation is good for all the studied models.
\newline\newline
Figure \ref{fig:z_components} shows $1+z_{\rm Sch}$ for three different versions of model 2. The contribution $1+z_{\rm Sch}$ fluctuates noticeably with peaks where the light rays come close to the central regions where the Schwarzschild mass is concentrated. The contribution to the total redshift is negligibly small in all three models though. The fluctuations in $1+z_{\rm Sch}$ is an order of magnitude larger in the model with small inner region which is due to the larger deviation of $A$ from 1 due to the higher density in the inner Schwarzschild region in this model. The accumulated $1+z_{\rm Sch}$ along the light rays is slightly smaller (approximately by a factor of $2.5$) in the model with large opaque centers which is as expected since the light rays in this model spend less time in a given Schwarzschild region which naturally makes the contribution to $1+z_{\rm Sch}$ from passing through any single Schwarzschild region smaller compared to the two other models where light rays can pass closer by the centers. This highlights the most important point with the redshift analysis (also discussed in \cite{Fleury}): The sole contribution to the redshift from passing through a Schwarzschild region (not including the boundary), is a {\em small} ISW effect which is negligible compared to the boundary and background contributions. Therefore, it is largely irrelevant whether or not there are sub-regions inside the Schwarzschild region which the light rays are not permitted to traverse. Consequently, the redshift will not in general be well described through a spatial average of the expansion rate as in equation \ref{eq:syksy_z} if light rays are restricted from propagating through certain spacetime regions and spatial averages are computed by omitting these regions. It must be emphasized that the analyses in \cite{av_obs1,av_obs2} do not specifically address this type of situation. The results obtained here show by explicit example that the results in \cite{av_obs1,av_obs2} cannot in general be naively extended to this type of situation where there are opaque regions in the considered spacetime.
\newline\newline
Another way of viewing the result from the above redshift analysis is that the main effect the Schwarzschild regions have on the redshift occur at the boundary and is thus independent of whether or not there are regions interior to the boundary that the light ray is not permitted to trace. It is then natural to consider if the result could be an artifact arising from the sharp boundary between the EdS and Schwarzschild regions. However, figure 3 in \cite{scSZ5} indicates\footnote{Specifically, the figure shows the contributions to the redshift along light rays in different LTB and Szekeres Swiss-cheese models. From the figure, it is seen that upon traversing an inhomogeneous region the redshift will, to a good approximation, be the same as if it had simply traveled in the background {\em regardless of the impact parameter}.} that the same can be expected in Szekeres \cite{Szekeres} and LTB \cite{LTB1,LTB2,LTB3} Swiss-cheese models, i.e. the redshift will be well described through the expansion rate averaged over all space, regardless of whether central regions are made opaque. This is despite the Szekeres and LTB models smoothly transitioning to their FLRW backgrounds rather than having abrupt $\delta$-function boundaries.
\newline\newline
For completeness, the redshift-distance relation is in figure \ref{fig:DA_shear} shown for model 2 with the interior region made transparent. In this case, all the methods for describing the mean observations predict the same result, namely that the mean redshift-distance relation is simply that given by the EdS background and as seen in the figure, this prediction is correct. Note though, that the dispersion around the mean is much larger here than in the corresponding model with opaque interiors. This is because the light rays are much more affected by shear in the case of transparent interiors.

\subsection{Redshift drift}
It was above found that the mean redshift is described well by the background EdS model and not the spatial average of the regions that the light rays were permitted to traverse. This could be interpreted as indicating that the redshift depends mostly on local quantities at the spacetime points of emission and observation. However, the relation between $k^{\mu}_0$ and $k^{\mu}_e$ depends on the light path and therefore the redshift must also to some extent depend on the particular light ray's path between the source and observer and not solely on the local qualities of spacetime at the points of emission and observation. This claim is in fact in agreement with the results found above, namely that the redshift in the Schwarzschild region is very different from the redshift in a pure Schwarzschild model while the redshift in the EdS region is only modified slightly compared to in a pure EdS spacetime - in accordance with the Schwarzschild region affecting $k^{\mu}$ very modestly compared to the EdS region.
\newline\indent
The redshift drift is more a local quantity than the redshift because the redshift drift depends on the derivatives $\frac{dk^T}{d\lambda}$ and not only fractions of $k^T$. Since $\frac{dk^T}{d\lambda}\neq k^Tk^T_{;T}$ in general and because $k^{\alpha}$ depends on an integral along the light ray, the redshift drift also depends to some extent on the spacetime along the individual light rays and not only on local quantities. It is therefore interesting to look at the mean redshift drift in exact, inhomogeneous solutions to the Einstein equations, in order to learn about the behavior of redshift drift.
\newline\newline
\begin{figure*}
	\centering
	\subfigure[]{
		\includegraphics[scale = 0.45]{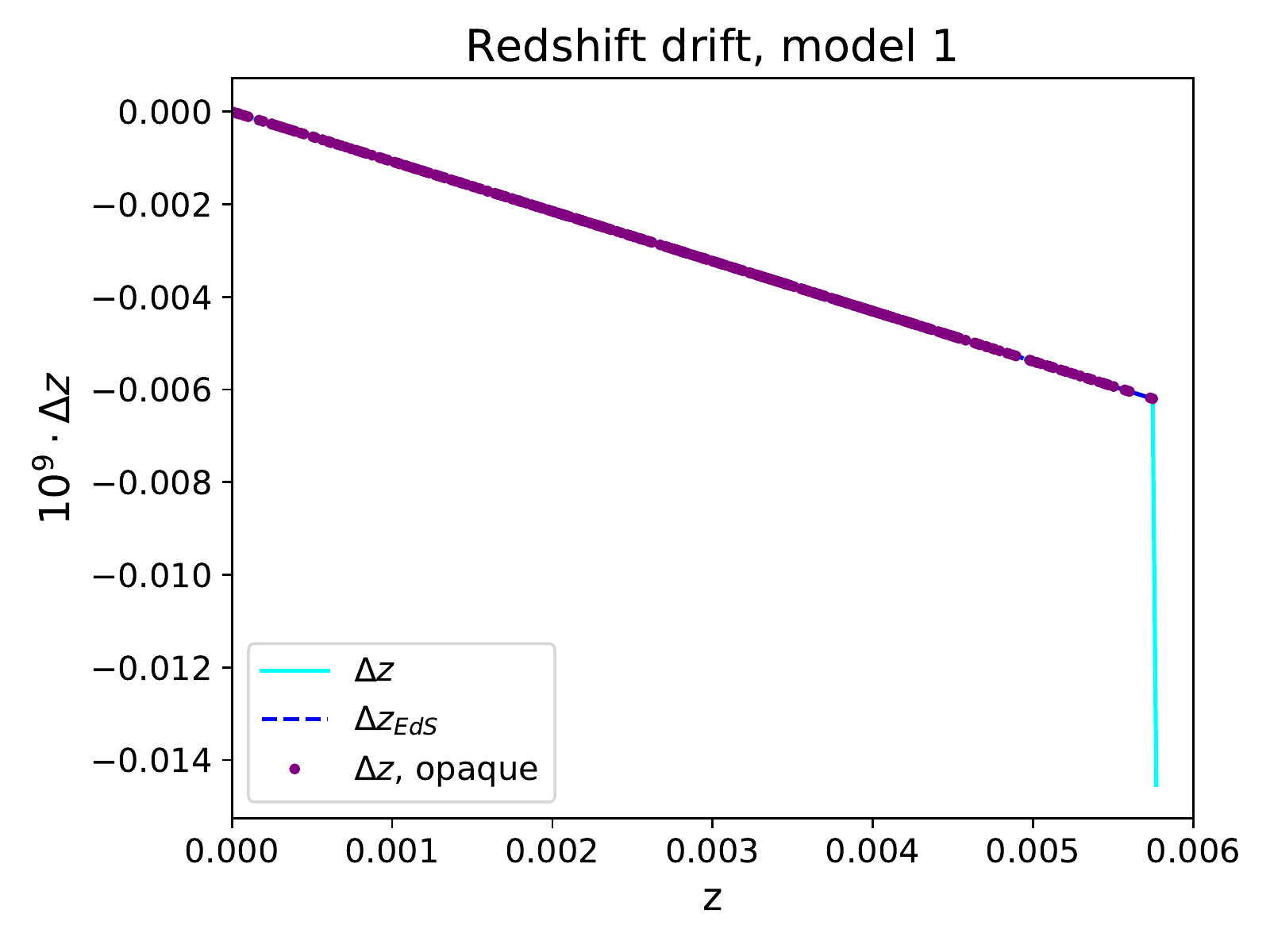}
	}
	\subfigure[]{
		\includegraphics[scale = 0.45]{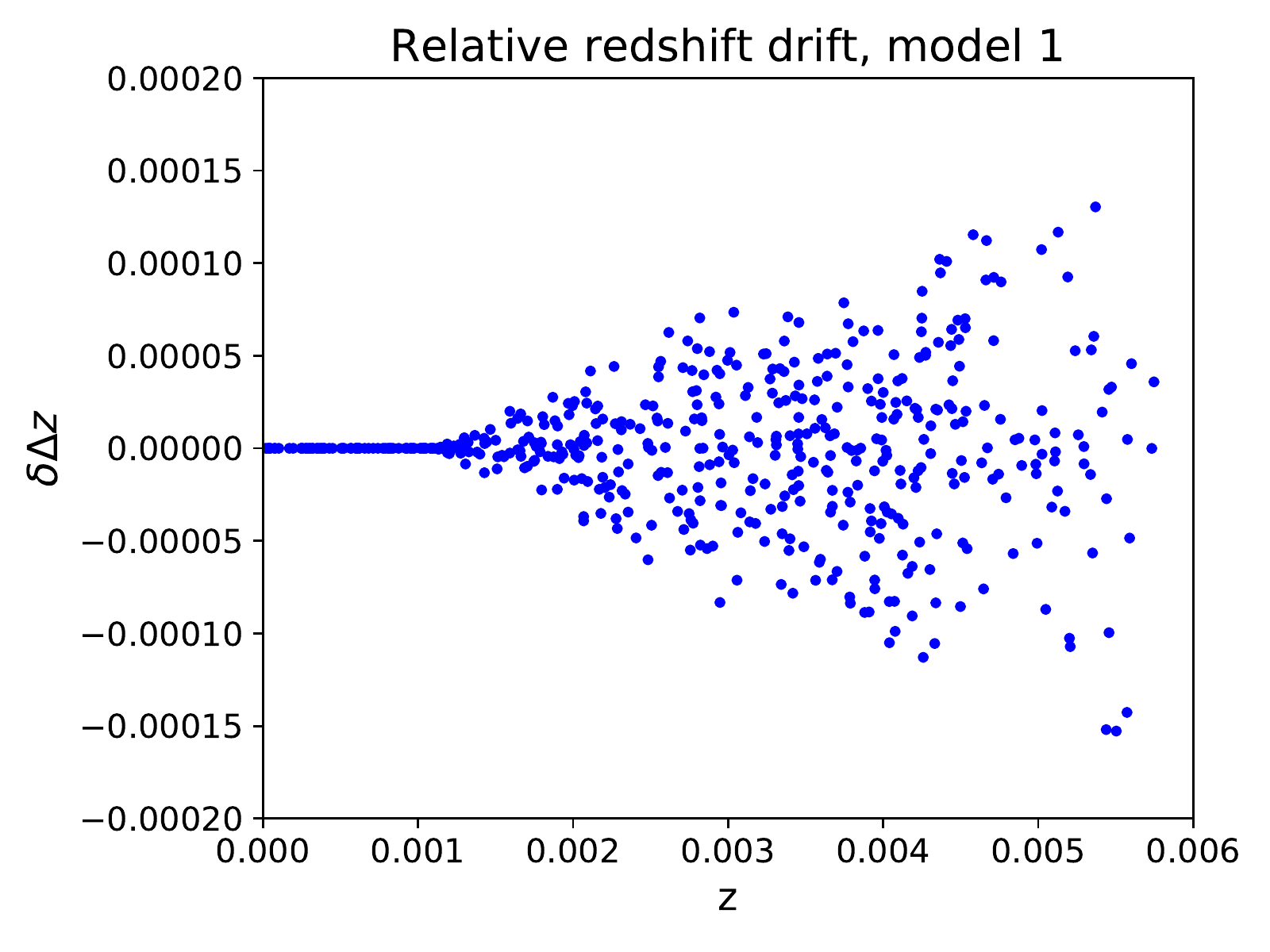}
	}\\
	\subfigure[]{
	\includegraphics[scale = 0.45]{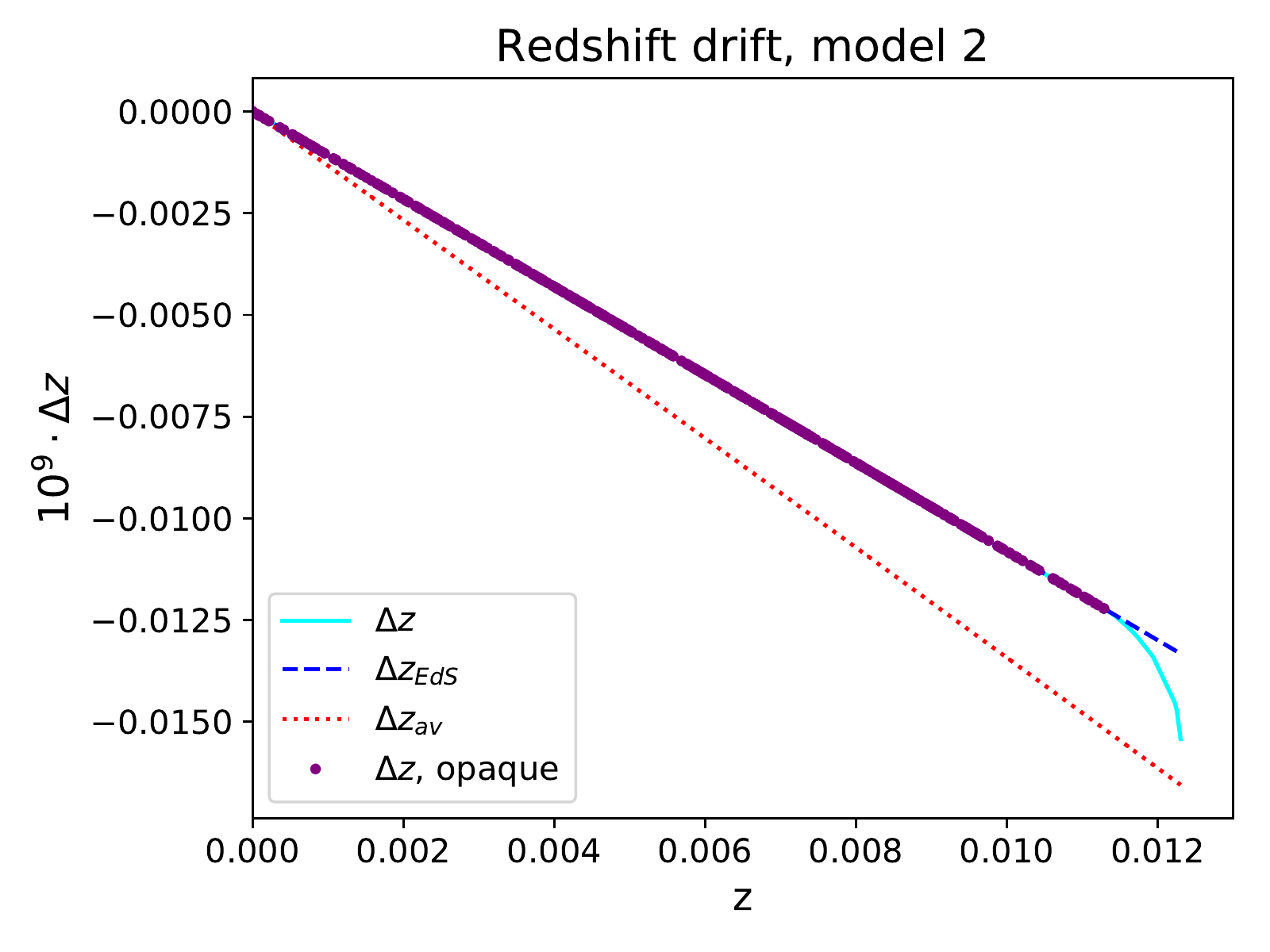}
}
\subfigure[]{
	\includegraphics[scale = 0.45]{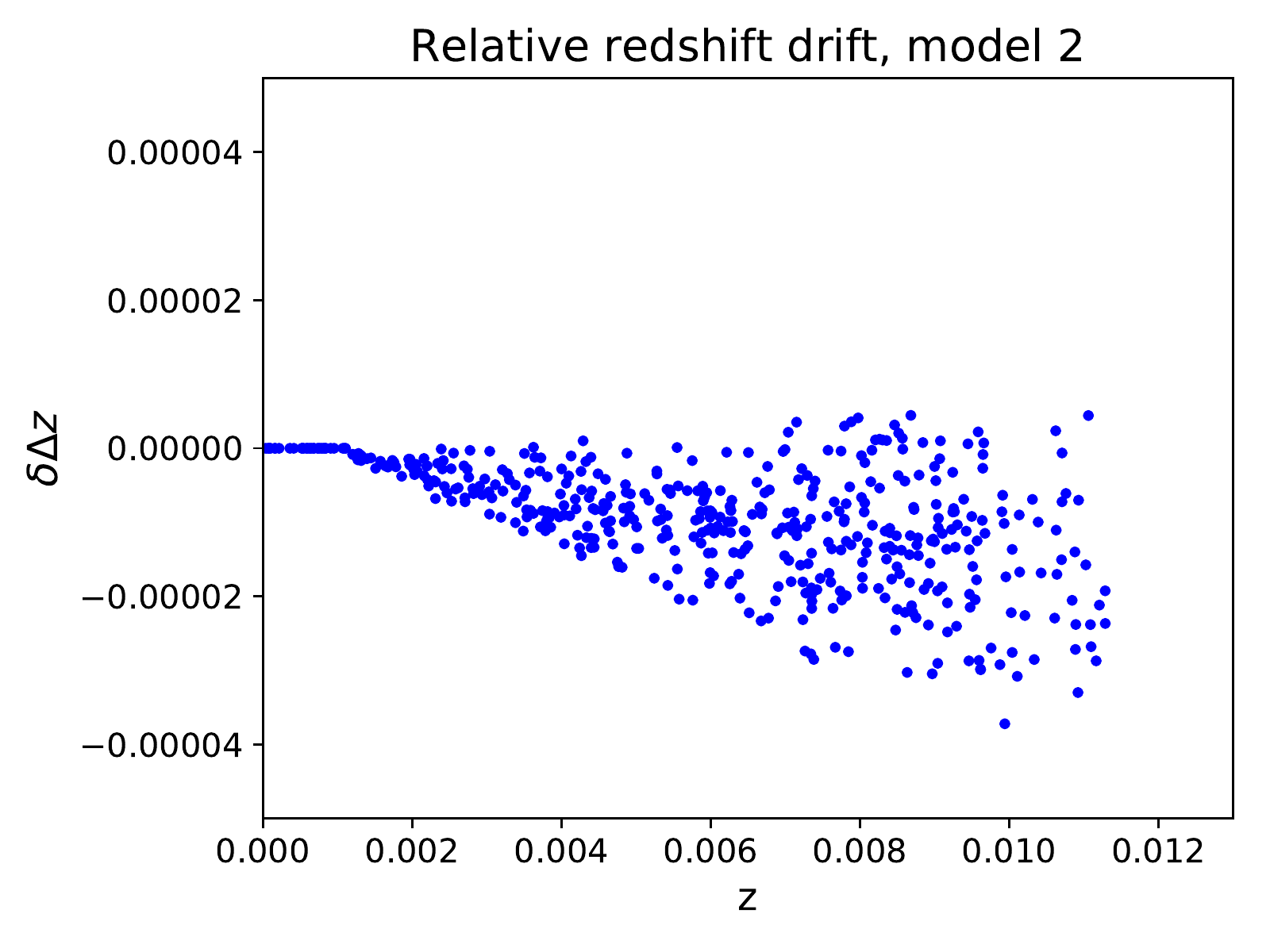}
}
	\caption{Redshift drift along light rays after traversing a single Schwarzschild region at different angles, with more radial impact parameters indicated by larger redshift. The top figures are for model 1 while the bottom figures are for the model 2. The figures to the left show the redshift drift for $\delta T_0 = 30$years, scaled by a factor of $10^9$. The thick solid lines (overlapping dots, really) represent light rays that have not entered the interior Schwarzschild region and are thus labeled as ``opaque''. The figures to the right show $\delta \Delta z:=\frac{\delta z - \delta z_{\rm EdS}}{\delta z_{\rm EdS}}$ and only depicts results for light rays that have not traveled through an interior Schwarzschild region.}
	\label{fig:zdrift}
\end{figure*}
Figure \ref{fig:zdrift} shows the redshift drift of 500 light rays after traversing a single Schwarzschild structure once with different impact parameters. The results are shown for models 1 and 2. The redshift drift is shown as a function of the value of the redshift after traversing the structure, where a larger redshift indicates a longer travel time and hence that the impact with the Schwarzschild metric has been with a more direct (radial) angle. The thick lines show the redshift drift for light rays that have not traveled through the interior regions and thus indicate the results for the opaque models. The thin cyan lines represent the models with transparent interiors and hence include light rays that have traveled through the interior regions. Clearly, the redshift drift is significantly altered along light rays that travel through the structures, but for the models with opaque regions, the redshift drift is modified very little compared to the EdS prediction. In fact, the dispersion seen in the subfigures to the right in figure \ref{fig:zdrift} is a manifestation of the precision ($10^{-16}$) of the computations being reached.
\newline\indent
Note that for model 1, $\Delta z_{\rm EdS}$ and $\Delta z_{\rm av}$ are identical and hence only $\Delta z_{\rm EdS}$ is shown. For model 2, the difference between the two is clear and $\Delta z_{\rm av}$ gives a poor description of $\Delta z$ (as expected considering the results found earlier regarding the mean redshift). For this reason, the quantity $\frac{\Delta z - \Delta z_{\rm av}}{\Delta z_{\rm av}}$ is not computed.
\newline\indent
The results presented in figure \ref{fig:zdrift} imply that the redshift drift will be well approximated by $\Delta z_{\rm EdS}$ in Einstein-Straus models if the interior regions are opaque. If on the other hand the interior regions are not opaque the mean will deviate from $\Delta z_{\rm EdS}$. However, the fraction of light rays that will traverse significantly into the interior Schwarzschild region will be modest so the bias will presumably not be very big. Indeed, the mean of the $\delta\Delta z$ for the results shown in figure \ref{fig:zdrift} for the two models with transparent interiors are approximately $0.0026$ ($ = 10^{15}M_{\odot}$) and $0.0011$ ($ = 10^{16}M_{\odot}$). For the opaque versions, the means are $5.5\cdot 10^{-7}$ and $9.5\cdot 10^{-6}$. The means of the opaque models are several orders of magnitudes smaller than for the transparent models but even for the transparent models, the means are quite modest (sub-percent). The accumulated effect of the structures on the redshift drift after traversing many structures will therefore presumably be modest despite the striking behavior for light rays traversing through the interiors.
\newline\indent
The prominent deviation between $\Delta z_{\rm EdS}$ and $\Delta z$ for light rays traversing the interior Schwarzschild regions is striking and merits further consideration. However, to meaningfully consider it further requires using model setups reminiscent of the relevant real scenario, i.e. light rays traversing through realistic regions with sources and observer position realistic compared to the sources expected to be used for redshift drift measurements in the future. This is beyond the scope of the work presented here. Note especially that the large deviation from $\Delta z_{\rm EdS}$ is due to $k^T_{;T}k^T$ beginning to deviate from $\frac{dK^T}{d\lambda}$. Since this apparently happens (significantly) only from propagating in the interior Schwarzschild region and not already when propagating through the exterior Schwarzschild region, the effect must depend crucially on the exact form of the metric functions. A meaningful quantification of the effect therefore requires using a realistic interior region (if it can be justified physically to let light rays travel through the interior regions at all).
 
\subsection{Note on interior FLRW regions}
As mentioned in section \ref{sec:modelSetup}, an FLRW metric can be joined with the Schwarzschild metric in such a way that the FLRW metric is interior to the vacuum Schwarzschild region. The boundary between the interior FLRW metric and the exterior Schwarzschild metric is analogous to the boundary between the exterior FLRW region and the exterior Schwarzschild region. Therefore, the redshift-distance results obtained above do not change much if an interior FLRW region is included. Specifically, the average expansion rate will still be that of the exterior FLRW region: Denoting the radial coordinate of the inner FLRW region by $\bar \xi$, the average expansion rate is
\begin{align}
\begin{split}
\left\langle \theta\right\rangle &=\frac{\left(\xi_0^3 - \xi_{\rm b}^3 \right)H_{\rm b} + \int_{\rm b} d\xi H_{\rm b} \delta(\xi-\xi_{\rm b})\xi^3 + \bar\xi_{\rm bie}^3H_{\rm bie} - \int_{\rm bie}d\bar\xi H_{\rm bie}\delta(\bar\xi - \bar \xi_{\rm bie})\bar \xi^3 }{\frac{1}{3}\xi_0^3}\\
& = \frac{\xi_0^3H_{\rm b}  + \bar\xi_{\rm bie}^3H_{\rm bie} - \bar\xi_{\rm bie}^3H_{\rm bie} }{\frac{1}{3}\xi_0^3} = 3H_{\rm b},
\end{split}
\end{align}
where subscripts $\rm b$ indicate FLRW values on the outer boundary, and $\rm bie$ on the inner boundary. Integrals $\int_{\xi_{\rm b}}$ are integrals over an interval that contains $\xi = \xi_{\rm b}$ and similarly for $\int_{\xi_{\rm bie}}$.
\newline\newline
The reason that the average expansion rate is still that of the outer FLRW region is that the expansion rate contribution from the inner FLRW region cancels with the contribution from the inner boundary. The same thing happens when computing the redshift: There is a non-negligible redshifting of light as it propagates through the inner  FLRW region, but this is largely canceled by the effects from the inner boundary (see the discussion around equations \ref{eq:z_boundary1}-\ref{eq:z_edS}).
\newline\newline
See e.g. \cite{Nottale1,Nottale2,Nottale3} for early studies on light propagation in models with both inner and outer FLRW regions joined through a Schwarzschild vacuum.

\section{Summary}\label{sec:Summary}
Several particular Einstein-Straus models were studied in terms of their light propagation qualities in order to compare mean observations with predictions based on spatial averaging and the Dyer-Roeder approximation. In agreement with earlier studies, it was shown with explicit examples that the Dyer-Roeder approximation gives a good description of the mean redshift-distance relation when contributions from shear is small. It was also shown that the relation based on spatial averages gives the same prediction as the Dyer-Roeder approximation in models with parameter values similar to those of earlier studies. When choosing a setup where the two approximations differ significantly, it was found that the Dyer-Roeder approximation gives the correct result while the method based on spatial averages underestimates the angular diameter distance significantly at high redshift. This shows that the relations obtained in \cite{av_obs1,av_obs2}, where the redshift-distance relation is based on spatially averaged quantities, cannot be naively generalized to spacetimes with opaque regions. While the results shown here are specifically based on Einstein-Straus models, results presented earlier (figure 3 in \cite{scSZ5}) indicate that this is the case for any Swiss-cheese model. It is important to learn how relevant these results are for real observations, i.e. which observables should be considered based on light rays propagating through a universe with opaque regions, what is the size and mass of these regions, and do the results presented here describe such opaque regions in terms of their effects on mean observations in general or are the results found here a specific feature of the Swiss-cheese models.

\section{Acknowledgments}
The author thanks Pierre Fleury for comments on the manuscript and explanations regarding results presented in \cite{Fleuryetal}. The work was completed using computer resources from the Finnish Grid and Cloud Infrastructure urn:nbn:fi:research-infras-2016072533.
\newline\newline
During the preparation of the original manuscript, the author was supported by the Independent Research Fund Denmark under grant number 7027-00019B.
\newline\newline
During the final stages of review, the author transitioned to being supported by the Carlsberg foundation. Final numerical checks made during this stage utilized UCloud services provided by SDU eScience Center.
\appendix
\section{Boundary contributions to expansion rate and shear}\label{app:Boundary}
This appendix serves to compute the contributions to the shear and expansion rate on the boundary between EdS and Schwarzschild regions by describing the region through a Lemaitre-Tolman-Bondi (LTB) model \cite{LTB1,LTB2,LTB3}.
\newline\newline
The LTB model  is a spherically symmetric dust solution to the Einstein equations. Its line element can be written as
\begin{equation}
	ds^2 = -dt^2 + \frac{R^2_{,\xi}(t,\xi)}{1-k(\xi)}d\xi^2 + R^2(t,\xi)d\Omega^2.
\end{equation}
The metric function $R$ can be considered an FLRW scale factor generalized to include radial dependence. This is seen through the dynamical equation $R_{,t}^2 = \frac{2M}{R} - k$.
\newline\newline
A straightforward computation shows that the local expansion rate and shear tensor components are given by
\begin{equation}
	\theta = 2\frac{R_{,t}}{R} + \frac{R_{,t\xi}}{R_{,\xi}}
\end{equation}
\begin{equation}
\begin{split}
	\sigma_{\alpha}^{\beta} = \text{diag}\left(0,\frac{2}{3},-\frac{1}{3},-\frac{1}{3} \right) \left( \frac{R_{,t\xi}}{R_{,\xi}} - \frac{R_{,t}}{R}\right)
\end{split}
\end{equation}
One may construct an LTB model consisting of a central static region surrounded by an EdS spacetime. The transition between the two regions can be constructed e.g. using a function that transitions smoothly between the two spacetimes. The transition between the two regions can be made as fast (in the radial direction) as one wishes, with the limit being a Heaviside function-like transition. This is the case that best corresponds to the situation in the Swiss cheese models studied in the main text, where the transition between EdS and Schwarzschild spacetimes occurs at a specific $\xi$-value rather than smoothly over an extended $\xi$-interval. In this case, $R_{,t}$ can be approximated as $\Theta(\xi_{\rm b})a^{EdS}_{,t}\xi$, where $\Theta(\xi_{\rm b})$ is the Heaviside function. The non-vanishing expansion rate of this model is
\begin{equation}
\begin{split}
	\theta(\xi\geq  \xi_{\rm b}) = 2\frac{a^{EdS}_{,t}}{a^{EdS}} + \frac{a^{EdS}_{,t} + \xi\delta(\xi-\xi_{\rm b})a^{EdS}_{,t}}{a^{EdS}} \\= 3H_{\rm EdS} + \xi\delta(\xi-\xi_{\rm b})H_{\rm EdS} ,
\end{split}
\end{equation}
where $\delta(\xi-\xi_{\rm b})$ is the delta function, entering into the expression as the derivative of the Heaviside function.
Similarly, the shear is
\begin{equation}
	\sigma_{\alpha}^{\beta} = \text{diag}\left(0,\frac{2}{3},-\frac{1}{3},-\frac{1}{3} \right)\xi H_{\rm EdS}\delta(\xi-\xi_{\rm b}).
\end{equation}

The boundary contributions (proportional to $\delta(\xi-\xi_{\rm b})$) can be used to estimate the boundary contributions of shear and expansion rate in the Swiss-cheese models studied in the main text.

\end{document}